\documentclass[%
reprint,
amsmath,amssymb,
aps,
]{revtex4-1}

\usepackage{graphicx}% Include figure files
\usepackage{dcolumn}% Align table columns on decimal point
\usepackage{bm}% bold math
\usepackage{bbm}% bbm fonts
\usepackage{mathrsfs}
\usepackage[table]{xcolor}
\usepackage{ulem} % strikethrough word

\DeclareMathOperator{\Li}{Li}
\DeclareMathOperator{\const}{const}
\DeclareMathOperator{\sign}{sign}
\begin{document}

	\preprint{APS/123-QED}
	
\title{\large\bf The higher-order magnetic skyrmions in non-uniform magnetic fields}

\author{M.\,S.\, Shustin$^1$}%
\email{mshustin@iph.krasn.ru}
\author{V.\,A.\, Stepanenko$^2$}%
\email{v-stepanen@mail.ru}
\author{D.\,M.\, Dzebisashvili$^1$}%
\email{ddm@iph.krasn.ru}

\affiliation{%
    $^1$Kirensky Institute of Physics, Federal Research Center KSC SB RAS, 660036 Krasnoyarsk, Russia\\
    $^2$Siberian Federal University, 660041, Krasnoyarsk, Russia}

\date{\today}% It is always \today, today,
%  but any date may be explicitly specified
	
	\begin{abstract}

For 2D Hubbard model with spin-orbit Rashba coupling in external magnetic field the structure of effective spin interactions is studied in the regime of strong electron correlations and at half-filling. It is shown that in the third order of perturbation theory, the scalar and vector chiral spin-spin interactions of the same order arise. The emergence of the latter is due to orbital effects of magnetic field. It is shown that for nonuniform fields, scalar chiral interaction can lead to stabilization of axially symmetric skyrmion states with arbitrary topological charges. Taking into account the hierarchy of effective spin interactions, an analytical theory on the optimal sizes of such states -- the higher-order magnetic skyrmions -- is developed for axially symmetric magnetic fields of the form $h(r) \sim r^{\beta}$ with \textcolor{black}{$\beta \in \mathbb{R}$}.

\begin{description}
	%\item[Usage]
	%Secondary publications and information retrieval purposes.
	\item[PACS number(s)]
	75.10.Hk, % Classical spin models)
	%02.30.−f, % Function theory, analysis
	%71.10.−w,  % Theories and models of many-electron systems
	71.27.+a,  % Strongly correlated electron systems; heavy fermions
	71.70.Ej,  % Spin-orbit coupling, Zeeman and Stark splitting, Jahn-Teller effect
	75.45.+j   % Macroscopic quantum phenomena in magnetic systems
	%75.75.−c  % Magnetic properties of nanostructures
	%\item[Structure]
	%You may use the \texttt{description} environment to structure your abstract;
	%use the optional argument of the \verb+\item+ command to give the category of each item.
\end{description}

\end{abstract}	
		
%\pacs{Valid PACS appear here}% PACS, the Physics and Astronomy
% Classification Scheme.
%\keywords{Suggested keywords}%Use showkeys class option if keyword
%display desired
\maketitle

%\tableofcontents

\section{\label{sec1}Introduction}

Starting with the pioneering papers of F. Bloch \cite{bloch30}, as well as L.D. Landau and E.M. Lifshitz \cite{landau35}, topological objects began to attract considerable attention in the physics of magnetism. Currently, topological objects in the physics of magnetic phenomena include both singular defects (domain walls, anisotropic two-dimensional vortices and Bloch points) and continuous objects, which include magnetic skyrmions (MS) \cite{skyrme61, skyrme62, bogdanov89, muhlbauer09}. The skyrmions were first studied by T. Skyrme in nuclear physics as topologically nontrivial configurations of the baryon field \cite{skyrme61, skyrme62}. Later, similar structures were predicted in magnetic systems \cite{bogdanov89} and experimentally detected in MnSi \cite{muhlbauer09}. In the last decade, the development of experimental technologies \cite{yang15, moreau16} has made it possible to create and study magnetic skyrmions in nanowires and thin magnetic films, inducing non-collinear magnetic structures in 1D and 2D systems.

Two-dimensional magnetic skyrmions are vortex-like distributions of magnetic moments in the plane $\mathbb{R}^2$. In the center of the vortex structure, the direction of the magnetic moment is opposite to the direction of the magnetic moments at the boundary of the skyrmion and outside its boundary. If the characteristic scales of a significant change in the magnetization distribution considerably exceed the interatomic distances, then it is possible to use a continuum approximation and consider the magnetization field ${\bf{m}}(\bf{r})$ as a smooth function of the spatial variable $\bf{r}$.

Practical interest in magnetic skyrmions is due to their topological stability. Therefore, the above mentioned vortex magnetic structures, despite the small (nano- or micrometer) scale, are stable against to defects and temperature fluctuations. Taking advantage of this stability, numerous schemes for using MS in logic devices \cite{zhang15, zavorka19} and memory \cite{yu17} are currently proposed. In particular, methods have been proposed for recording and reading magnetic information by creating and moving a spin-polarized MS current along one-dimensional paths \cite{yu17}. An important point of such constructions was the assumption that the topology of the MS profiles does not change when writing and reading information.

Mathematically, the question of continuous deformation of magnetic structures is related to the homotopy theory \cite{shvartc17}. Two magnetization configurations are called topologically (homotopically) equivalent if there is a way of their continuous deformation into each other without overcoming an infinite energy barrier. And conversely, two configurations are topologically non-equivalent if such continuous deformation is impossible.

Usually 2D MS are considered as \textcolor{black}{smooth magnetic textures} on a two-dimensional real plane with $\bf{m}\in\mathbb{S}^2$, $\bf{r}\in \mathbb{R}^2$. Since the two-dimensional plane $\mathbb{R}^2$ is also one-pointwise compactified into the sphere $\mathbb{S}^2$, various magnetic configurations are characterized by the mapping $\mathbb{S}^2\to\mathbb{S}^2$, and the homotopy group of such configurations is the homotopy group $\pi_{2}\left(\mathbb{S}^2\right) \sim\mathbb{Z}$. The latter is isomorphic to the group of integers $\mathbb{Z}$ and nontrivial magnetic configurations can be characterized by such non-zero numbers.

To establish an unambiguous correspondence of the magnetic configuration $\bf{m}(\bf{r})$ to the elements of the homotopy group $\pi_{2}\left(\mathbb{S}^2\right)$, the concept of the mapping degree $Q$ is used. The latter is also sometimes called "topological index" or "topological charge". For the mapping  $\mathbb{S}^2\to\mathbb{S}^2$, it has the form \cite{shvartc17}:
\begin{equation}
\label{Q}
Q = \frac{1}{4\pi}\int_{-\infty}^{\infty}\int_{-\infty}^{\infty}
\left( {\bf{m}} \cdot \left[\,\frac{\partial {\bf{m}}}{\partial x}\times \frac{\partial {\bf{m}}}{\partial y} \,\right] \right)\,dx\wedge dy,
\end{equation}
where $x$ and $y$ are the plane coordinates in $\mathbb{R}^2$.
This characteristic takes integer values $Q\in\mathbb{Z}$, which show how many times during mapping $\mathbb{S}^2 \to \mathbb{S}^2$ the vector $\bf{m}$ sweeps full solid angle $4\pi$.

Until recently, the vast majority of 2D MS studies were related to axially symmetric structures with $|Q|=1$. This is due to the fact that it is precisely such axially symmetric structures that are stabilized by Dzyaloshinsky-Moriya (DM) interaction, which is most actively involved for MS modeling. In recent studies, it has been predicted that frustrated exchange interaction in combination with DM interaction can lead to stabilization of axially symmetric higher-order magnetic skyrmions (HOMS) with $|Q|>1$ \cite{leonov15, ozawa17, rozsa17}. Also, on the basis of numerical modeling, the existence of skyrmion states of nontrivial morphology with arbitrary values of topological indices \cite{foster19, rybakov19, kuchkin20}, which were called skyrmion bags, was predicted. \textcolor{black}{Recently, skyrmion bags have been experimentally obtained in the interior of the thin plate of the B20-type FeGe chiral magnet, as an 2D constituent parts of 3D magnetic structures, having high topological charges and called as skyrmion tubes \cite{tang21}. However, to date, both skyrmion bags and HOMS have been studied extremely limitedly.}

Meanwhile, the search for axially symmetric MS with arbitrary topological indexes is of interest from the point of view of creating an element base of both classical and quantum computing devices. For example, the implementation of high-$Q$ magnetic states can significantly increase the density of recorded classical information by increasing the space of read states \cite{yu17}. From a perspective of quantum computing devices, MS with large $|Q|$ may be of interest as promising objects for the implementation and management of Majorana modes (MM) \cite{zlotnikov21, gungordu22}. Particularly, in a recent paper it was shown that in order to implement MM on the HOMS, it is necessary that the topological charge  of the
latter was even \cite{yang16}. Since such states have not been discovered experimentally  the search for MM was carried out on bound states of MS and superconducting vortex \cite{rex19}. The stability of such a bound state depends on many factors and may be due to the action of stray fields of the vortex on the MS \cite{dahir19, menezes19, dahir20, andriyakhina21, andriyakhina22, apostoloff22}. In this regard, the further search for implementation of axially symmetric HOMS in non-uniform magnetic fields is relevant and of practical interest.

In this paper, stabilization of the HOMS due to the three-spin interaction, induced by strong electron correlations and inhomogeneous magnetic field, is predicted and the characteristic sizes of such structures are analyzed.
In Sec.\ref{sec2} for two-dimensional strongly correlated ensemble of electrons with Rashba spin-orbit interaction (SOI) on a triangular lattice, the effective spin-spin interactions are analyzed in the third order of perturbation theory.
Both symmetric and chiral magnetic interactions of different nature are derived and a hierarchy of amplitudes of such interactions is established. In Sec.\,\ref{sec3} taking into account non-uniform magnetic field the phenomenological high-spin Heisenberg model with such interactions is formulated. Further, in Sec.\,\ref{sec4} it is shown that when only orbital effects of magnetic field are taking into account, the formulated model makes it very easy to describe the formation of the HOMS with arbitrary $Q$.
In Sec.\,\ref{sec5} the Zeeman effects of magnetic field on HOMS's sizes are studied and analytical description of the HOMS is carried out for \textcolor{black}{the case of linear increased} inhomogeneous magnetic field. In the Sec.\,\ref{sec6} the robustness of HOMS against to field profiles variation is demonstrated. Prospects for the detection and practical use of HOMS in non-uniform fields are discussed in Sec.\,\ref{sec7}.  Sec.\,\ref{sec8} describes the main results of the study.

\section{\label{sec2} Chiral interactions in the Hubbard model with Rasba spin-orbit coupling in magnetic field }

%Chiral interactions in an strongly correlated system

Let us consider the issue of formation and competition of chiral interactions of different nature in a 2D ensemble of strongly correlated electrons. As a starting point for such consideration, we take the Hubbard model \cite{hubbard65, valkov01} on the two-dimensional lattice subjected to an external magnetic field and with taking into account the Rashba SOI \cite{banerjee14, malki20}. Its Hamiltonian has the form:
\begin{eqnarray}\label{Hub_mod}
\mathcal{H}&=& \mathcal{H}_0 + \mathcal{V},\\
\mathcal{H}_0 &=& -\sum_{f\sigma}\left(\mu +
\textcolor{black}{g\sigma\mu_B B_f}
 \right)c^+_{f\sigma}c_{f\sigma}+U\sum_f n_{f\uparrow}n_{f\downarrow},\nonumber\\
\mathcal{V} &=& \sum_{\langle fg \,\rangle \sigma}\left(  t_{fg}\,c^+_{f\sigma}c_{g\sigma}+h.c.\right) + \nonumber\\
&+&i\alpha\sum_{\langle fg\rangle\sigma\sigma'} \left({\bf{d}}_{fg}\times{\bf{\tau}}_{\sigma\sigma'} \right)_{z}c^{+}_{f\sigma}c_{g\sigma'}.\nonumber
\end{eqnarray}
Here $c_{f\sigma}(c^{+}_{f\sigma})$ are the electron annihilation (creation) operators at the site $f$ with the spin projection on the quantization axis ($z$-axis) $\sigma=\pm 1/2$, $\mu$-- the chemical potential, the bare on-site energy of an electron is set to zero, \textcolor{black}{$B_f$\,-- external magnetic field in the cite $f$}, $n_{f\sigma}=c^{+}_{f\sigma}c_{f\sigma}$-- the operator of electron number at the site $f$ and with spin projection $\sigma$, $U$-- the energy of Coulomb repulsion of two electrons on one site which in the regime of strong electron correlations is supposed to be the largest parameter of the model. In the magnetic field the hopping integral $t_{fg}$ between sites $f$ and $g$ due to Peierls substitution \cite{peierls33, landau10} acquires exponential factor:
$$t_{fg}\to t_{fg}\exp\left(\frac{ie}{c\hbar}\int_{r_s=f}^{r_f=g}\bf{A}\cdot d\,\bf{l}\,\right),~~{\bf{l}} = \left(x,\,y\right),$$
where $e$ -- electron charge, $c$ -- speed of light, $\hbar$ -- Planck's constant and integration of the 1-form constructed on the magnetic field vector potential $\bf{A}$ is carried out along the straight line connecting the sites $f$ and $g$
in the direction of the unit vector ${\bf{d}}_{fg}$
pointing from $f$ to $g$.
Angle brackets under the sum symbol in (\ref{Hub_mod}) indicate that hoppings only between the nearest neighbors with an amplitude $t_{fg}=t$ are taken into account.
The last term in the expression (\ref{Hub_mod}) stands for the Rashba SOI with intensity $\alpha$.
The components of the vector $\bf\tau$ are the three Pauli matrices $\tau^x$, $\tau^y$ and $\tau^z$, and
the subscript at the right parenthesis means $z$-component of the corresponding vector.

The presence of SOI suggests fabrication of a two-dimensional heterostructure.
Candidate materials whose electronic properties can be described by this model include cuprate superconductors, rare-earth intermetallides, and water-intercallated sodium cobaltites
$\textrm{Na}_{x}\textrm{CoO}_{2} \cdot y\textrm{H}_{2}\textrm{O}$ above the superconducting transition temperature $T_c=5$ K \cite{kanigel04, zheng06, baskaran03, ogatta03, kumar05, valkov15, valkov16, valkov19}.
Specificity of the electronic structure of such materials is due to the fact that dynamics of the current carriers is  predominantly in (quasi-)two-dimensional layers. Apart from that, such two-dimensional layers should be brought into contact with another two-dimensional material and the system should be placed in external magnetic field.

Consider the system (\ref{Hub_mod}) at  half filling, assuming that on average there is exactly one electron at each lattice site.
The structure of effective interactions dependent on the spin degrees of freedom of the model (\ref{Hub_mod}) was obtained using the method of unitary transformationsins in the Hilbert space of many-body states \cite{bir72}.
A detailed derivation of these expressions in the case of $U\gg t,\alpha,h$ is given in Appendices A and B.
We will consider the regime of sufficiently weak SOI and Zeeman splitting: $U\gg t\gg \alpha, h$. To be more specific, it will be also assumed that: $t/U\sim\alpha/t\sim \varepsilon$, where $\varepsilon$ is a value of the first order of smallness.
Then, up to and including the third order of $\varepsilon$, we obtain the following three effective spin-spin interactions:
\begin{eqnarray}\label{HJ}
\mathcal{H}_{J}&=&-\sum_{[fg]}\mathcal{J}\cdot{\bf{S}}_f\cdot{\bf{S}}_g ,\\ \label{HD}
\mathcal{H}_{D}&=&\sum_{[fg]}\mathcal{\bf{D}}_{fg}\cdot\left[{\bf{S}}_{f} \times {\bf{S}}_g\right],\\ \label{HK}
\mathcal{H}_{K}&=&\sum_{[fgl]\in \Delta}\mathcal{K}\cdot{\bf{S}}_f\cdot\left[{\bf{S}}_g\times {\bf{S}}_l \right],
\end{eqnarray}
where ${\bf S}_f$ is a vector spin operator on the site $f$, \textcolor{black}{and notation $[\,f\,g\,l\,]\in\Delta$ in Eq.\,(\ref{HK}) means that the three nearest sites $f$, $g$ and $l$ form elementry triangular plaquet and the summation is carried out over this plaquettes.}

The indirect exchange interaction (\ref{HJ}) is of antiferromagnetic (AFM) type with  exchange integral $\mathcal{J} = -4t^2/U$.
Square brackets under the sum symbols mean that each pair or triple of sites is counted once.
Dzyaloshinsky-Moriya (or vector chiral) interaction (\ref{HD}) with the amplitude $\mathcal{D}=|\mathcal{\bf{D}}_{fg}|=8t\alpha/U$ formaly appears in the second order of pertubation theory (see Eq. (\ref{Heff2_tot})). But since $\alpha\sim t\varepsilon$ the amplitude $\mathcal{D}$ is actually the value of the third order of smallness.

Operator (\ref{HK}) stands for the three-spin (or scalar chiral) interaction.
With regard to this interaction it should be noted that in the nearest neighbor approximation it is equal to zero identically for a square lattice due to the homeopolarity condition.
Therefore, it is convenient to consider triangular lattice.
In this case the amplitude of the three-spin interactions, according to the Appendix B, is: $\mathcal{K} =24\cdot (t^3/U^2)\cdot\sin(\pi\Phi_{\Delta})$.
If the magnetic flux $\Phi_{\Delta}$ (given in units of magnetic flax quantum $\phi_{0} = ch/2e$)  through elementry triangular plaquet $\Delta$, is taken from the interval: $1/6\pi \lesssim \Phi_\Delta\ll 1$, then the amplitudes  $\mathcal{K}$ and $\mathcal{D}$ are of the same order: $t\varepsilon^2$ (see Appendix B).
Moreover, for HOMS, the contribution to the energy functional from the three-spin interactions, as we will see below (see Eq. (\ref{EF2})), is proportional to the value of $|Q|>1$, which enhances the magnitude of this interaction as compared to the DM one.
Therefore, both chiral interactions (DM and three-spin) should be considered jointly.

 Note, that the expressions (\ref{HD}) and (\ref{HK}) for DM and three-spin interactions have been derived separately out of the Hubbard model (see, for example \cite{malki20, banerjee14, sen95, motrunich06, bauer14} and the cited literature). In the present paper, these expressions were obtained jointly within unified approach (see Appendices A and B) making it possible to establish a hierarchy between effective parameters (\ref{JDKA})
as well as to establish the microscopic nature of the scalar and vector chiral interactions. In the next section, we will phenomenologically formulate the classical Heisenberg model with scalar and vector chiralities, keeping in mind the parameters hierarchy (\ref{JDKA}) as well as a specific dependence of $\mathcal{K}$ on magnetic field.

\section{\label{sec3} The High-Spin Heisenberg model with a scalar and vector chiralities}

Having derived the microscopic model with symmetric (\ref{HJ}) and two chiral interactions (\ref{HD}) and (\ref{HK}) let us phenomenologically reformulate it in the classical (high-$S$) limit. As a justification for this, we note that a similar but much more cumbersome microscopic derivation of effective spin-spin interactions, can be carried out for the case $S_f>1/2$. In particular, chiral interactions of the described type can be obtained within the framework of the $s-d(f)$--exchange model for  arbitrary value of the spin $S$ \cite{banerjee14, komarov17, villalba22}.
On the other hand, it is known that taking into account the direct exchange interaction, as well as the multi-orbitality of magnetoactive ions, can lead to the  effective ferromagnetic (FM) exchange interaction in the Hubbard multi-orbital model \cite{irkhin93, gavrichkov17}. Thus we assume that the exchange integral $\mathcal{J}$ in (\ref{HJ}) is positive.
\textcolor{black}{Furthermore, in magnetic materials there is almost always a natural magnetic crystallographic anisotropy. Sometimes the anisotropy can reach quite large values comparable with the exchange interactions \cite{coulon06, bogani08, zhang13}.}

\textcolor{black}{Thus, along with the interactions (\ref{HJ})-(\ref{HK}), we have to include in our effective model  an anisotropy term and also take into account Zeeman term wich in fact  strictly follows from the original Hubbard hamiltonian (\ref{Hub_mod}) if magnetic field is turned on}
\begin{eqnarray}\label{HZ}
 \mathcal{H}_{A}=-\mathcal{A}\sum_f\left(S^z_f\right)^2,~~\mathcal{H}_{Z}&=& -g \mu_B \sum_{f}B_f\,S^z_f.
\end{eqnarray}
If $\mathcal{A}>0$ ($\mathcal{A}<0$), then the anisotropy has an easy-axis (easy-plane) character. The need for accounting for the terms (\ref{HZ}) is motivated by the fact that if we consider only orbital effects of the magnetic field then to stabilize HOMS the anisotropy should be involved. Moreover, the anisotropy term
naturally arises in the continuum description of antiferromagnetic skyrmions \cite{potkina20}, where Zeeman splitting results in renormalization of the single-ion anisotropy constant  $\mathcal{A}$.

Thus the phenomenological spin Hamiltonian
which we are going to study has the form:
\begin{eqnarray} \label{H_micr}
\mathcal{H}_{eff}=\mathcal{H}_{J}+\mathcal{H}_{D}+\mathcal{H}_K+\mathcal{H}_{Z}+\mathcal{H}_A
\end{eqnarray}
with hierarchy of energy parameters obtained in the previous section:
 \begin{equation}\label{JDKA}
\mathcal{J} \gg \mathcal{D} \sim \mathcal{K} \gg  \mathcal{A},~g\mu_B B.
\end{equation}

With account for the classical limit, as well as a large spatial scale of MS as compared to the interatomic spacing, the model (\ref{H_micr}) can be considered as continuous. Hence, introducing the spin vector field $\bf{m}(\bf{r})$ with the norm $|{\bf m}({\bf r})|=1$, in the usual way we obtain an expression for the energy of continuum version of the model (\ref{H_micr}) on triangular lattice:
\begin{eqnarray}
\label{H_cont}
E = E_{J}+E_{D}+E_{K}+E_{Z}+E_{A},
\end{eqnarray}
with
\begin{eqnarray}
E_{J}&=&\frac{\sqrt{3}S^2\mathcal{J}}{2}\int \sum_{\mu=x,y,z}\left( \bf{\nabla} m_\mu\right)^2 ds, \\
E_{D}&=&\frac{\sqrt{3}S^2\mathcal{D}}{a} \int\left(m_z\cdot\left(\bf{\nabla} \bf{m}\right)- \left({\bf{m}} \bf{\nabla}\right)\cdot m_z\right)ds,~~~~~\\
E_{K}&=&\int 2S^3\mathcal{K}(x,y)\cdot\left( {\bf{m}} \cdot \left[\frac{\partial {\bf{m}}}{\partial x}\times \frac{\partial {\bf{m}}}{\partial y}\right] \right)ds,\\ \label{EZ}
E_{Z}&=&\int \frac{2Sg\mu_B}{a^2\sqrt{3}} ~B(x,y)\cdot(1-m_z)ds,\\ \label{EA}
E_{A}&=&\frac{2S^2\mathcal{A}}{a^2\sqrt{3}}\int (1-m_z^2)ds,
\end{eqnarray}
where ${\bf{\nabla}}=\left(\partial/\partial x, \partial/\partial y,0\right)$, $ds = dx\wedge dy$, $a\,-$ triangular lattice parameter and the integration is carried out in the $\mathbb{R}^2$ domain, where magnetic moments exist.

It is important to note that expressions (\ref{H_cont})-(\ref{EA}) describe the excitation energy of the system over homogeneous state with $m_{z}({\bf r})\equiv 1$. It can be easily  seen that the term $E_{K}$, characterizing the scalar chiral interaction, is proportional to the integral over $\mathbb{R}^2$ of the density of the topological charge (\ref{Q}) with the kernel
\begin{equation}\label{Hb}
2S^3\mathcal{K}(x,y)=\frac{48\cdot t^3}{U^2}S^3\sin\left(\pi\Phi_\Delta(x,y)\right), \end{equation}
describing the spatial profile of the externally  applied magnetic  field $B(x,y)$. In the case of homogeneous field the contribution of $E_K$ is reduced to a shift of the energy $E$ by the value proportional to the topological charge of the magnetic configuration, and therefore cannot give rise a new magnetic structure.
Below we will show that due to  inhomogeneity of the external magnetic field the scalar chiral interaction can lead to stabilization of both simple MS with $n=1$ and HOMS with $n>1$.

In the Sec.\ref{sec3} and \ref{sec4} we will demonstrate the formation of HOMS in the external magnetic field
\begin{eqnarray}\label{H_lin}
B(r)&=&
B_0 \cdot h(r),
\end{eqnarray}
described by the simplest axially symmetric linear dependence $$h(r)=r,$$
whith $r$ being the distance from the center of the skyrmion measured in units of interatomic spacing $a$:  $r=|{\bf{r}}|/a$. Parameter $B_0$ in (\ref{H_lin}) is induction of the magnetic field at the distance $a$ from the center of MS.

At the same time, due to smallness of $\Phi_{\Delta}(r)$, it can be assumed that the spatial profile of the amplitude $\mathcal{K}(r)$ (\ref{Hb}) follows the external magnetic field profile (\ref{H_lin}):
\begin{eqnarray}\label{K_lin}
\mathcal{K}(r)\sim h(r),
\end{eqnarray}
and therefore can be varied experimentally. Later, in Sec.\,\ref{sec5} we will consider a more general profile function, $h(r) = r^{\beta}$ (\textcolor{black}{$\beta \in \mathbb{R}$}), in order to consider the robustness of HOMS against magnetic field variation.

It is important to stress that in (\ref{EZ}), (\ref{H_lin}) and below we \textcolor{black}{will mainly} consider only the z-component of the magnetic field neglecting its radial components: ${\bf{B}} = (\,0,\, 0,\,B(r))$. In this case, to satisfy the Maxwell 's equation ${\bf{\nabla}}\cdot{\bf{B}}=0$, it is necessary to assume that $\partial{B}\,/\,\partial z = 0$. Thus, \textcolor{black}{
except the end of Sec.\,\ref{sec6}}, we will consider idealistic case of non-uniform magnetic field  ${\bf{B}}$ with the profile $h(r) =r^{\beta}$, which is independent of the coordinate $z$.

\section{\label{sec4} HOMS in the chiral Heisenberg model without Zeeman splitting}

Let us consider the issue of HOMS stabilization within the chiral Heisenberg model (\ref{H_micr}), which takes into account the interaction of orbital degrees of freedom with an external magnetic field ($K\neq 0$), but neglects the Zeeman splitting ($H=0$ see Eq. (\ref{en_params})). In particular, such a situation can be realized \textcolor{black}{both for FM skyrmions at zero values of the $g-$factor and for AFM skyrmions \cite{potkina20}.}

\begin{figure}[htb!]\center
\includegraphics[width=0.42\textwidth]{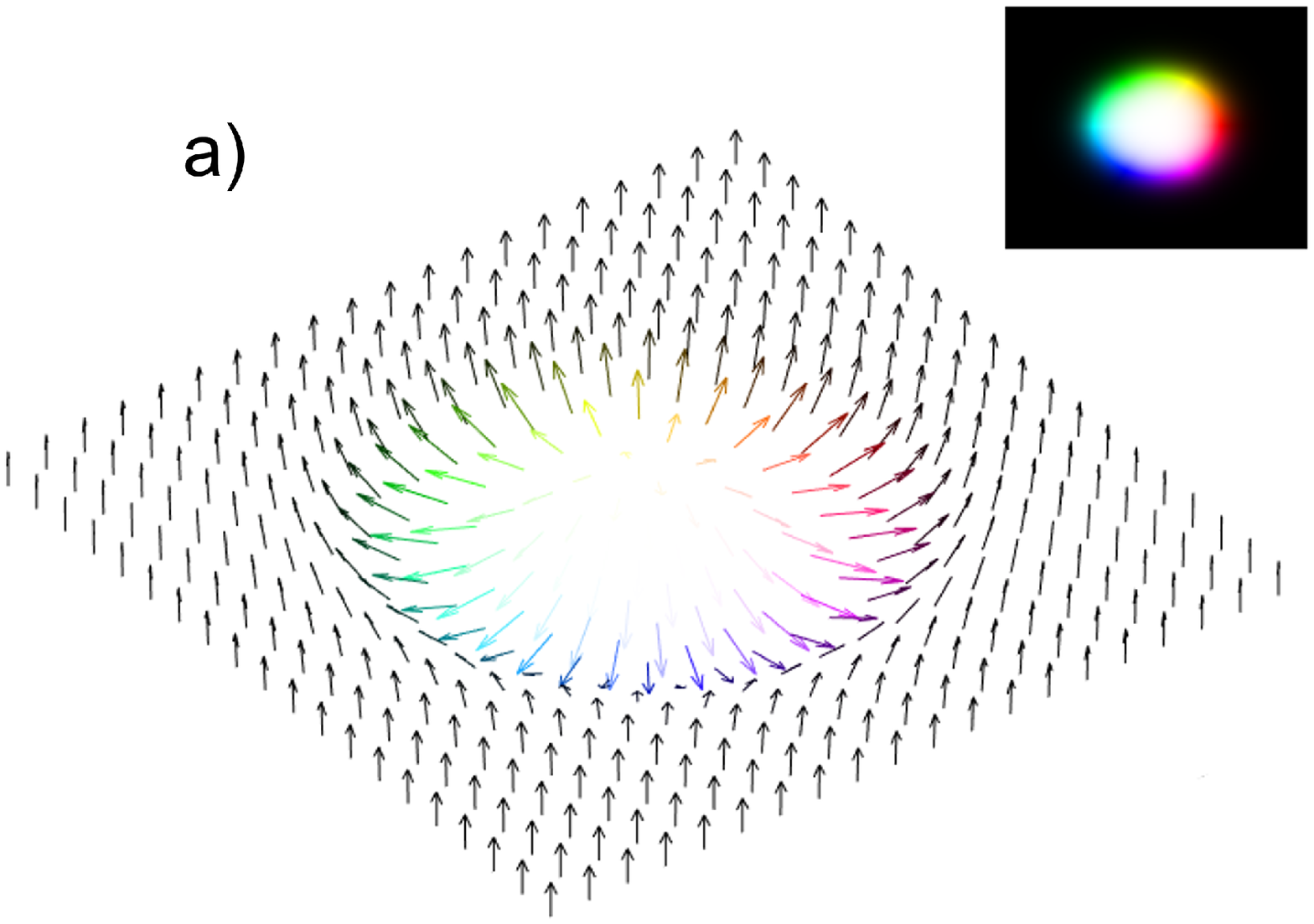}
\includegraphics[width=0.42\textwidth]{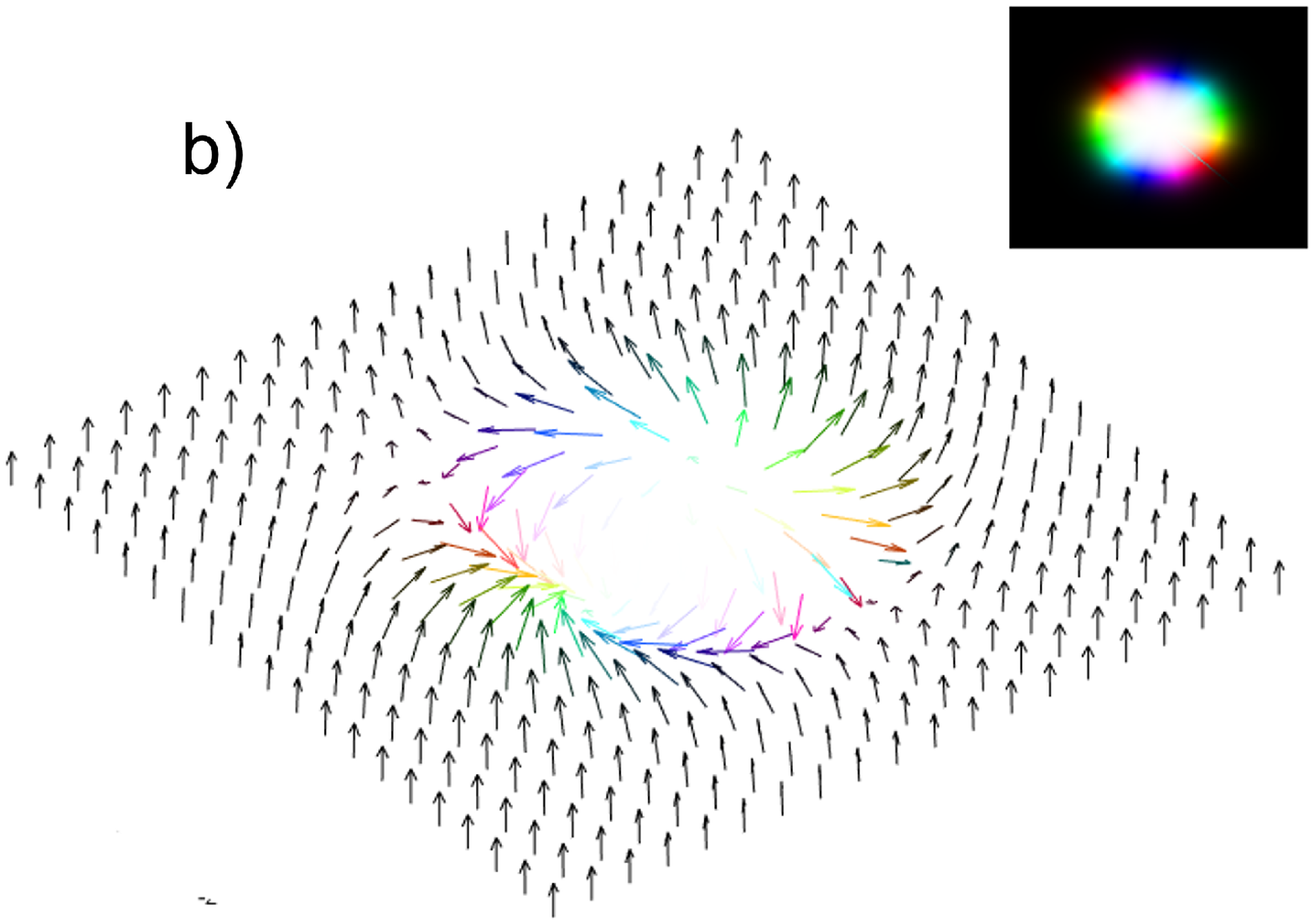}
\caption{\label{Fig1}
Spatial profiles of HOMS: a) $n=1$, b) $n=2$. The inserts on the right visualize the profiles by means of color scheme used below in the text. The black and white color correspond to the directions with $m_z=+1$ and $m_z=-1$, respectively.
In the case of $m_z\neq 1$, the color corresponds to the direction of projection of the magnetization field $\bf{m}(\bf{r})$ onto the plane $\textrm{XoY}$. If: $\{m_x,m_y\}=\{1,0\}$, then the color is red; $\{m_x,m_y\}=\{\cos \frac{2\pi}{3},\sin \frac{2\pi}{3}\}$ -- green; $\{m_x,m_y\}=\{\cos \frac{4\pi}{3},\sin \frac{4\pi}{3}\}$ -- blue. For clarity, vector profiles are presented in the same color code.}
\end{figure}

To study the conditions of HOMS stabilization we will use the variational approach for the classical continuum functional (\ref{H_cont}) corresponding to the effective Hamiltonian (\ref{H_micr}) with the hierarchy of parameters set above in (\ref{JDKA}). Minimization of the functional (\ref{H_cont}) will be carried out with the vector profile of the HOMS defined by the following parametrization \cite{wang18}:
\begin{eqnarray}
\label{sk_param}
&~&m_x = \sin \Theta \cos n\varphi ;~~
m_y = \sin \Theta \sin n\varphi ;~~
m_z = \cos \Theta; \nonumber\\
&~&\Theta(r,\,R,\,w) = 2\arctan{\left(\frac{\cosh\left(R/w \right)}{\sinh\left(r / w \right)} \right)},
\end{eqnarray}
where $\phi$ is the polar coordinate of the vector ${\bf r}\in\mathbb{R}^2$, $\Theta$ and $n\phi$ are polar and azimuthal angles of the unit vector $\bf{m}(\bf{r})$. The corresponding vector profiles (with square grid for simplicity) are shown in Fig.\ref{Fig1}.

\textcolor{black}{
The parametrization with the radial function $\Theta(r)$ is widely used to describe skyrmions with $n = |Q| = 1$ and has demonstrated excellent agreement with physical and numerical experiments \cite{wang18}. In the case of HOMS, we have neither numerical nor physical results for the similar comparison. Moreover, a detailed numerical analysis of the $\Theta(r)$ behavior for various parameters and boundary conditions is rather hard and beyond the scope of this work. Thus, let's present in our study only the qualitative arguments, which are usually used in justifying the ansatz (\ref{sk_param}) for MS with $n=1$, and demonstrate that they are also valid for HOMS in nonuniform fields.}

\begin{figure}[htb!]\center
\includegraphics[width=0.42\textwidth]{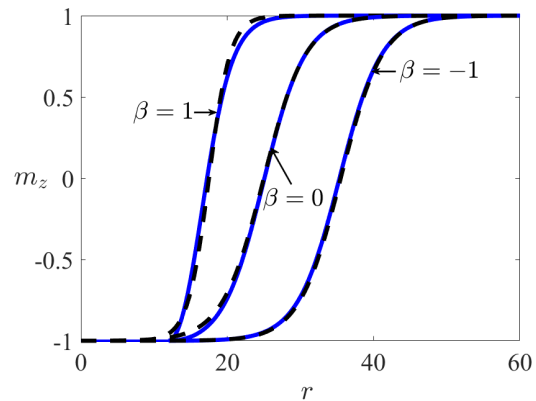}
\caption{\label{Fig2}
\textcolor{black}{Radial profiles of domain walls. Solid lines describe numerical exact solutions of Eq.(\ref{1D_eq}) with $J=20$, $A=0.6$, $H=0.05$, $h(r)=r^{\beta}$. The dashed lines are calculated using the ansatz (\ref{sk_param}) with the fitting parameters $\{R,w\}$: $\{17.5, 2.9\}$ for $\beta=1$; $\{25, 5.7\}$ for $\beta = 0$; $\{35.5, 5.7\}$ for $\beta=-1$.}}
\end{figure}

\textcolor{black}{Having parameterized the HOMS by Eq.(\ref{sk_param}) we, in fact, considered it as an axially symmetric 1D $\pi$-domain wall with $w$ being its width, and $R$ -- the distance from the center of the skyrmion to the domain wall middle with both defined in units of $a$. The equation for the spatial profile of such a domain wall can be obtained by considering the 1D restriction of the functional (\ref{H_cont})-(\ref{EA}) and writing the Euler-Lagrange equations for it \cite{braun94}. Assuming that $m_z = \cos\Theta$, considering the radial direction and neglecting the magnetization changing in the azimuthal direction in the film plane we obtain following equation:
\begin{equation}\label{1D_eq}
\frac{d^2\,\Theta}{d\,r^2} - \frac{A}{2J}\sin2\Theta -\frac{H}{2J}\,h(r)\,\sin\Theta = 0.
\end{equation}
It is easy to verify that the function (\ref{sk_param}) is a solution of (\ref{1D_eq}) at $H=0$ and $R \gg w$. If $H \ll J$, the exact solutions of Eq.\,(\ref{1D_eq}) can still be approximated with a high precision by the anzats (\ref{sk_param}), as shown in Fig.\,\ref{Fig2} for field profile $h(r) = r^{\beta}$ at $\beta = 0,\,\pm1$. The deformation of the domain wall profile due to non-uniform magnetic field can be significant if $H \cdot h(\tilde{r}) \sim J$ in the interval $R-2w<r<R+2w$. However, in what follows we will consider regimes with $H \cdot h(r) \ll J$ and $w \ll R$ neglecting the deformation of the domain wall radial profile. In other words, we will consider the function $\Theta(r)$, Eq.(\ref{sk_param}), as an appropriate anzats for our study.}

\textcolor{black}{Note that in the presented qualitative consideration, neither the scalar chiral interaction, $K$, nor vorticity, $n$, appears in the Eq.(\ref{1D_eq}). Thus, the arguments presented here to justify the validiry of the 2$\pi$-domain wall ansatz are equally valid for both ordinary skyrmions and HOMS. Also note that at the end of Sec.\,\ref{sec6} we present another numerical justification of the ansatz (\ref{sk_param}) for the case of linearly decreased magnetic fields.}

Suppose that the center of the skyrmion core (point $r=0$) coincides with the projection of the axis of rotation of the axially symmetric profile of the applied magnetic field
$B(r)$ onto the plane $\mathbb{R}^2$ hosting magnetic moments. Then, substituting the parametrization (\ref{sk_param}) into the classical continuum version (\ref{H_cont}) of the Hamiltonian (\ref{H_micr}), we obtain an expression for the energy functional allowing for magnetic skyrmion solutions:
\begin{eqnarray}  \label{EF2}
{E} = E_{J}+E_{D}+E_{K}+E_{A},
\end{eqnarray}
where
\begin{eqnarray}  \label{EJ2}
E_{J}&=& \frac{J}{2} \int_0^{\infty} \left[\left(\frac{d\Theta}{dr}\right)^2+\frac{n^2}{r^2}\sin^2\Theta\right]r\,dr,\\
E_{D}&=&\delta_{n,1}\cdot \frac{D}{\pi} \int_0^{\infty} \left[\frac{d\Theta}{dr}+\frac{\sin 2\Theta}{2r}\right]r\,dr, \\
E_{K}&=&\frac{K\,n}{2}\int_0^{\infty}h(r)~\sin \Theta ~ \frac{d\Theta}{dr}\,dr, \\
E_{A}&=&\frac{A}{2}\int_0^{\infty} \sin^2 \Theta ~ r\,dr. \label{EA2}
\end{eqnarray}
Here the following energy parameters were introduced:
\begin{multline}\label{en_params}
J = 8\pi\sqrt{3}\,S^2\,\frac{t^2}{U};~
D = 16\pi^2\sqrt{3}\,S^2\,\frac{t\alpha}{U};~A = \frac{8\pi}{\sqrt{3}}\,S^2\,\mathcal{A};\\
K = 192\, \pi^2\,S^3\,\Phi^{(0)}_{\Delta}\,\frac{t^3}{U^2};~H = \frac{32\,\pi}{3}S\,\Phi^{(0)}_{\Delta}\, g\, \mu_B\, \bar{B},
\end{multline}
where $\bar{B} = \phi_0\,/\,a^2$, $\Phi^{(0)}_{\Delta}$ -- magnetic flux of the field $B_0$ through a triangular plaquette measured in units of magnetic flux quantum $\phi_0= ch\,/\,2e$. Here we have also introduced the parameter $H$ describing the energy of the Zeeman splitting, which will be taken into account in the next section, see Eq.\,(\ref{E_gen2}).
Hereinafter we will assume that HOMS are characterized by a narrow domain wall.
This means that the parameter $\rho = R/w \gg 1$. Then, calculating integrals (\ref{EJ2})-(\ref{EA2}) under these assumptions and using approximations made in \cite{wang18} (see Appendix D), we obtain expression for the functional $E(n,\rho, w)$:
\begin{eqnarray}
\label{E_gen}
E = J\left(\rho + n^2\rho^{-1}\right) - D\,\rho\,w\,\delta_{n,1} - K\,n \,\rho\,w + A\,\rho\,w^2.\nonumber\\
\end{eqnarray}

It can be seen that in the case of $n=1$, the scalar and vector chiral interactions either enhance or compete with each other. The functional of type (\ref{E_gen}) for $n=1$ was considered in \cite{wang18}. Making substitution $D\to D+K$ in their result we find the optimal sizes of a MS in our model:\begin{eqnarray}
\label{Rw_wang_H0}
&~&R_* = (D+K)\sqrt{\frac{J}{4JA^2 - (D+K)^2A}}\,;~~w_* = \frac{D+K}{2A},\nonumber\\
&~&n=1.
\end{eqnarray}
Note that for competing scalar and vector chiral interactions, $\sign(K)=-\sign(D)$, the expressions (\ref{Rw_wang_H0}) are still correct if the conditions we need $R_*\gg w_ *\gg $1 are fulfilled.

\begin{figure}[htb!]\center
\includegraphics[width=0.5\textwidth]{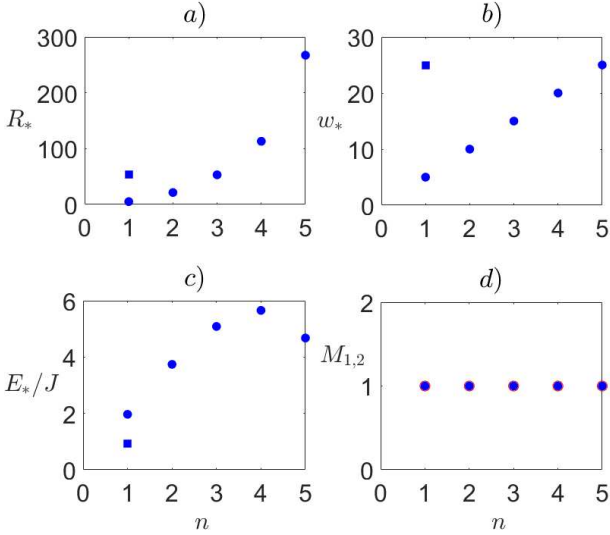} \caption{\label{2} Dependences of the optimal parameters of HOMS: radius $R_*$, domain wall width $w_*$ and energy $E_*$ on the topological charge $n$. \textcolor{black}{The circles correspond to the energy parameters in eV: $J=160$, $D=0$, $K=2$, $A = 0.2$. For the case $n=1$, the squares also represent the results of the calculation if the Dzyaloshinskii-Moriya interaction appear, $D = 8$\,eV}. The numbers $M_1$ and $M_2$ from Eq.\,(\ref{S12}) are shown in plot d) as blue dots and red circles, respectively. These dependencies correspond to the minimum of the energy functional (\ref{E_gen}).}
\end{figure}

\begin{figure}[htb!]\center
\includegraphics[width=0.23\textwidth]{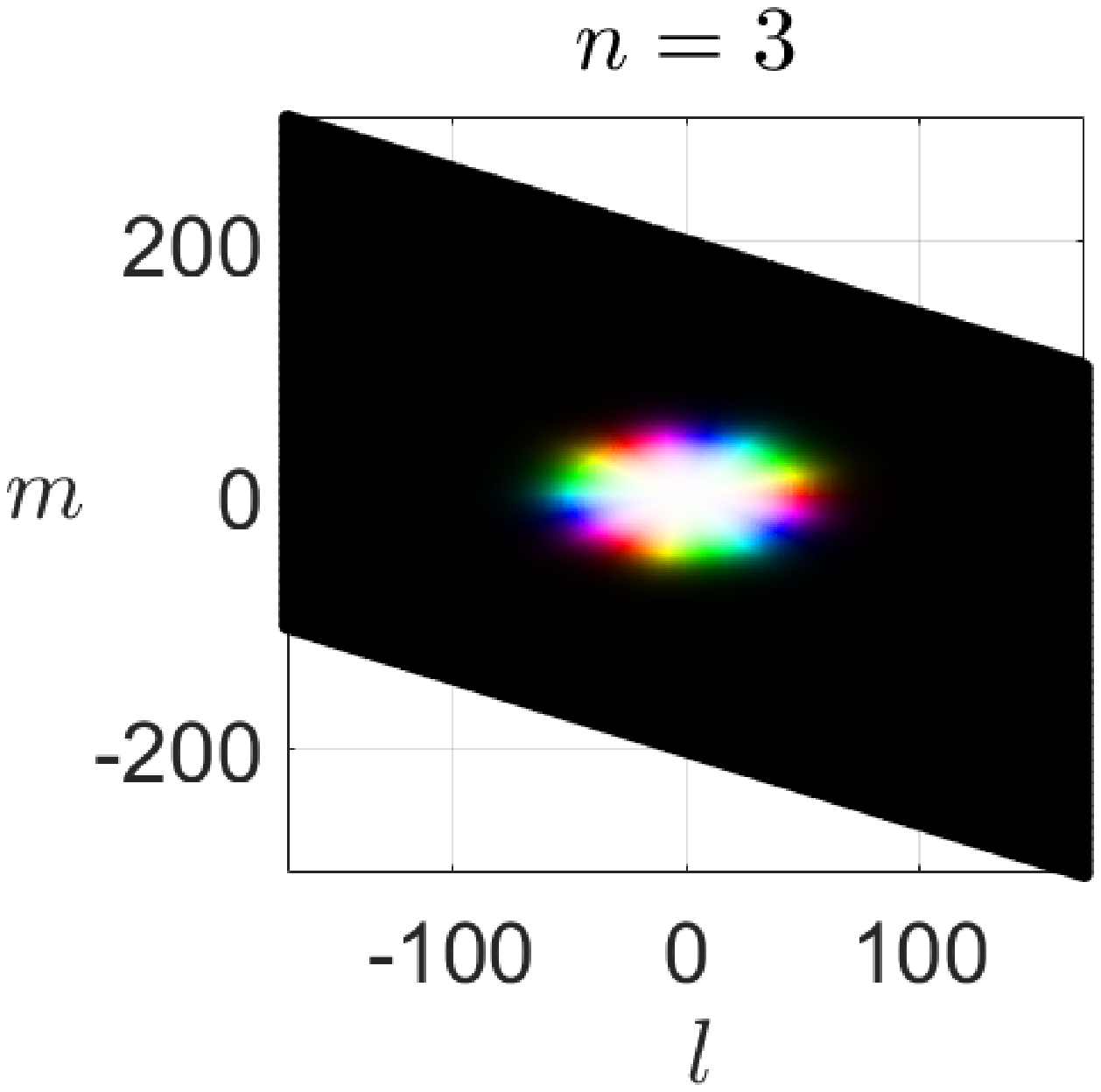}
\includegraphics[width=0.23\textwidth]{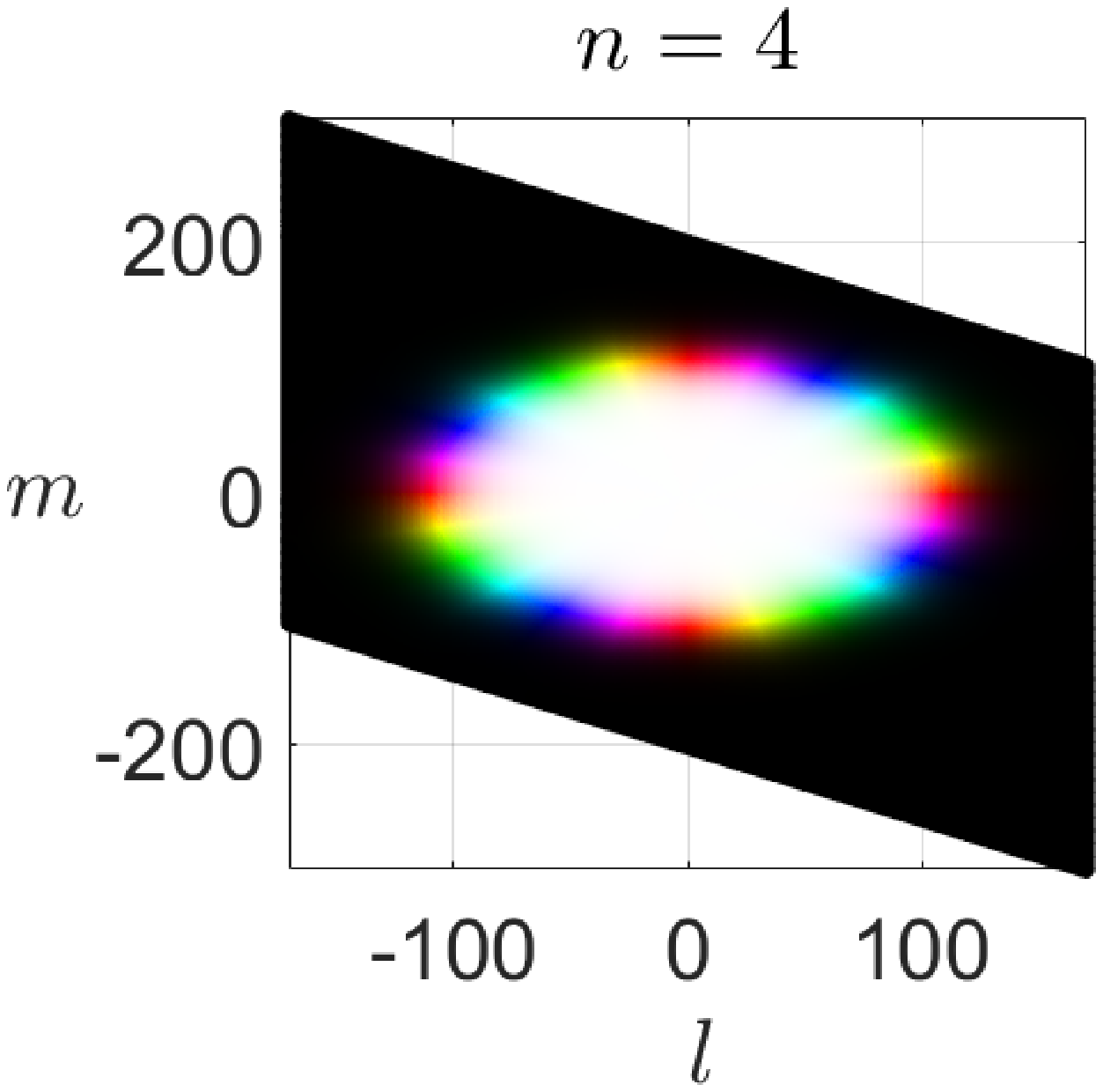}
\caption{\label{3} Spatial profiles of the HOMS with $n=3,4$ with the optimal sizes $R_*$ and $w_*$ corresponding to the Fig.\,\ref{2}}.
\end{figure}

In the case of the HOMS with $n>1$, the optimal parameters $R_*$ and $w_*$ can also be found simply by solving a system of algebraic equations
\begin{eqnarray}
\label{Rw_sms_ddm}
J(\rho^2 - n^2)-Kn\rho^2w+A\rho^2w^2  = 0;\nonumber\\
-Kn\rho^2w + 2A\rho^2w^2 = 0.
\end{eqnarray}
Since the parameter $D$ is absent here, the only factor of stabilization of the HOMS is the scalar chiral interaction. The solution of the Eq.\,(\ref{Rw_sms_ddm}) is convenient to write in variables:
\begin{equation}\label{def_nwR}
n_* \in \mathbb{N};~w_s = K\,/\,2A;~R_s = w_s\sqrt{1 - {K^2}/{4JA}}.
\end{equation}
where
$$n_*=\textrm{round}\left[\,\sqrt{\frac{1/2}{1-\left({w_s}/{R_s}\right)^2}}\,\right]=\textrm{round}\left[\,\sqrt{\frac{2\,J\,A}{K^2}}\,\right],$$
and $\textrm{round}(\ldots)$ is rounding to the nearest integer. The parameters $R_s$ and $w_s$ determine the MS optimal size in the uniform field at $n=1$ and $D\to K$.
In the variables introduced, the solution of the Eq.\,(\ref{Rw_sms_ddm}) has the form:
\begin{eqnarray}
\label{toy_model}
&~&w_*(n)=w_s\,n,\nonumber\\
&~&R_*(n)=\sqrt{2}\,w_s\,n^2\,\left(2n_*^2-n^2\right)^{-1/2},\nonumber\\
&~&E_*(n)=\sqrt{2}\,J\,\left(\frac{n}{n_*}\right)\left(2n_*^2-n^2\right)^{1/2}.
\end{eqnarray}
As $n$ increases from 2 to maximally allowed $n_{max}=\sqrt{2}n_*$, the HOMS domain wall width $w_*$ grows linearly, while the radius $R_*$ grows quadratically (see Fig.\,\ref{2}). Hence, skyrmions become larger and have a sharper domain wall as $n$ increases (see Fig.\,\ref{3}). At the same time, the obtained hierarchy of parameters $J \gg |K| \gg |A|$ leads to the desired properties of HOMS: $n_*$, $\rho_*$, $w_* \gg 1$.

Note that under the conditions
$$\frac{K\,n}{2A} \to \pm\sqrt{\frac{J}{A}}$$
the HOMS sizes (both $R_*$ and $w_*$) tend to infinity. The divergence of the radius $R_{*}$ is evident from the Eq.\,(\ref{toy_model}). The divergence of $w_*$ can be seen by solving the system (\ref{Rw_sms_ddm}) with respect to $w$ and estimating the convergence conditions of solutions according to the method \cite{cherepanskiy20}. This feature of the model is due to the absence of the Zeeman term in (\ref{E_gen}).

Note that the property $w>0$ leads to the condition $\sign(K)=\sign(A)$ (see Eq.\,(\ref{def_nwR})). So, for an easy-axis (easy-plane) magnets, the stabilization of the HOMS is possible when a magnetic flux is positive (negative) at the skyrmion core.

The topological charge of the HOMS is defined as
$$Q = \frac{n}{2}\big(\, \cos \Theta|_{r=\infty}-\cos \Theta|_{r=0} \,\big)=n.$$
According to the homotopy theory, the continuous deformation of the between HOMS with different $n$ is impossible without overcoming the infinite energy barrier. Accounting for the discreteness of the lattice makes the barriers finite and leads to the possibility of tunneling the system between metastable states with different $n$. Calculation of the minimum energies of such transitions and lifetimes of metastable states is beyond the scope of this study.

In order to explicitly demonstrate that the found HOMS states are metastable ones we will calculate the characteristics
\begin{eqnarray}\label{S12}
    {M}_{1} &=& \left.\sign\left(\,\frac{\partial^2 E}{\partial w^2}\,\right)\right|_{\genfrac{}{}{0pt}{}{w=w_*}{R=R_*}},\\
    {M}_{2} &=& \left.\sign\left(\frac{\partial^2 E}{\partial w^2}\cdot\frac{\partial^2 E}{\partial R^2}-\left(\frac{\partial^2 E}{\partial w \, \partial R}\right)^2\right) \right|_{\genfrac{}{}{0pt}{}{w=w_*}{R=R_*}}.\nonumber
\end{eqnarray}
Then the condition $M_{1,2}=+1$ corresponds to the local minimum of the functional $E$ (\ref{E_gen}).

Thus, the formulated model (\ref{H_micr}) demonstrates the possibility of stabilization the HOMS with $|Q|>1$ due to the scalar chiral interaction and orbital effects of magnetic field but in the absence of Zeeman ones. Wherein,
there are some constraints on the model parameters  for which the formation of HOMS is possible. Below, we will demonstrate the effects induced by Zeeman splitting.

\section{\label{sec5} The Effect of Zeeman Spliting}

Taking into account the interaction of the magnetic field (\ref{H_lin}) with spin degrees of freedom, the Zeeman term  $E_Z$ must be added to the energy functional (\ref{EF2}):
\begin{eqnarray}
\label{E_gen2}
E &\to& E + E_Z,\nonumber\\
E_Z &=& \frac{H}{2}\,\int_0^{\infty} h(r)\,(\,1-\cos \Theta\,)\,r\,dr = \nonumber\\
&=& H\left(\frac{w}{2}\right)^{\beta + 2}\Gamma_{\beta+2}\,\Li_{\beta+2}\left(-e^{2\rho},\,\textcolor{black}{\delta r}\,\right).
\end{eqnarray}
\textcolor{black}{Here we assume that $h(r) = 0$ if $r<\delta\,r$ where parameter $\delta\,r \geq 0$ depend on
$\beta$ and is introduced to regularize the integral (\ref{E_gen2}). Under this assumption, in Eq.\,(\ref{E_gen2}) the incomplete polylogarithm is naturally arised
\begin{eqnarray}
\label{polylog_inc}
\Li_{\beta+2}\left(-e^{2\rho},\,\delta r\, \right)=\frac{-1}{\Gamma_{\beta+2}}\int_{\delta r}^{\infty}\,\frac{t^{\beta+1}\,dt}{e^{t-2\rho}+1},
\end{eqnarray}
where $\Gamma_{\beta+2}$ -- gamma function of argument $\beta+2$. It can be seen from the consideration of Eq.\,(\ref{polylog_inc}) that it is necessary to assume $\delta r>0$ if $\beta < -1$ and we can assume $\delta r=0$ otherwise. In what follows, we will consider in detail the field profiles with $\beta \geq -1$ and therefore will take $\delta r = 0$. In this case we will use the asymptotics
\begin{eqnarray}
\label{polylog}
&~&\left.\Li_{\beta+2}\left(-e^{2\rho},\,\delta r=0\, \right)\right|_{\rho \gg 1}=\\
&=&\sum_{2k \leq \beta + 2}(-1)^k\left(1- 2^{1-2k} \right)\frac{\left(2\pi\right)^{2k}}{2k\,!}\frac{B_{2k}}{\Gamma_{s+1-2k}}\left(2\rho\right)^{s-2k},\nonumber
\end{eqnarray}}
where $B_{2k}$ is the Bernoulli number.

The Eq.\,(\ref{E_gen2}) describes the Zeeman splitting energy of the HOMS in magnetic field with the profile $h(r) = r^{\beta}$. For the simple "linear" case, $\beta=1$, the asymptotics of Zeeman term is:
\begin{eqnarray}
\label{wang}
E_Z = \bar{H}\left(\,\rho^3 w^3 + c^2\rho w^3 \,\right),
\end{eqnarray}
where $\bar{H}=H/3$, $c=\pi/2$. Then the excitation energy of the HOMS with an arbitrary $n$ has the form
\begin{eqnarray}
\label{E_gen_al1}
E &=& J\left(\rho + n^2\rho^{-1}\right) - Kn\rho w + A\rho w^2 + \bar{H}w^3\left(\rho^3 + c^2\rho \right).\nonumber\\
\end{eqnarray}
Their sizes are defined by the system of two equations
\begin{eqnarray}
\label{Rw_sms_ddm2}
&~&J(\rho^2 - n^2)-Kn\rho^2w+A\rho^2w^2 + \bar{H}w^3\left(3\rho^4 + c^2\rho^2 \right) = 0;\nonumber\\
&~&-Kn\rho^2w + 2A\rho^2w^2 + 3\bar{H} w^3\left(\rho^4 + c^2\rho^2 \right)=0.
\end{eqnarray}
that can be represented as Eq. of the 5th degree in $w$:
\begin{eqnarray}
\label{w5_sms_ddm}
&~&w^5\,6\bar{H}^2\,c^4+w^4\,7A\,\bar{H}\,c^2+ w^3\,\left(2A^2-2Kn\,\bar{H}\,c^2\right)+\nonumber\\
&+&w^2\,\left[-A\,Kn-3J\,\bar{H}\left(n^2+c^2\right) \right]-w\,2J\,A+J\,Kn=0,\nonumber\\
\end{eqnarray}
when taking into account the relations
\begin{eqnarray}\label{rho_vs_w}
R &=&\rho\,w =  \sqrt{\frac{Kn-2A\,w-3\bar{H}\,c^2\,w^2}{H}}=\\
&~&~~~~~~~~~~~~=\frac{\pi}{2}\sqrt{-\left(w - w_+\right)\left(w - w_-\right)}\,;\nonumber\\
w_{\pm}&=&\pm\frac{1}{H\,c^2}\left(\mathcal{R}\mp A\,\sign(H)\right);~~\mathcal{R}=\sqrt{A^2+K\,n\,H\,c^2}.\nonumber
\end{eqnarray}
It was found that out of the five roots of this Eq.\,(\ref{w5_sms_ddm}), there is always only one actual root that satisfies the conditions: $\rho,\,w\, \in \mathbb{R}$ and $\rho,\,w\,\gg \,1$. Moreover, from Eqs.\,(\ref{w5_sms_ddm}) and (\ref{rho_vs_w}) it can be seen that HOMS sizes are restricted by the inequalities
$$1 \ll w_* \leq w_{+};~~1 \ll R_* \leq 2\,\mathcal{R}\,/\,\pi\,|H|.$$
Thus taking into account the Zeeman splitting leads to a principal limitation on the skyrmion sizes,
which are obtained from Eq. (\ref{w5_sms_ddm}) at $A=0$
(see Appendix \ref{appendix_D})
\begin{align}\label{Rw_an_H}
w_* \approx \frac{1}{\sqrt{H}}\,y - \frac{\pi^2}{12\,JH}\,y^4+\frac{5\pi^2\,\sqrt{H}}{288\left(JH\right)^2}\,y^7\,, \\
R_* \approx \sqrt{{K\,n}\,/\,{H}-\pi^2 \, w_*^2\,/\,4}\,,~~y = \sqrt{\frac{K\,n}{n^2 + c^2}}\,. \nonumber
\end{align}
At the same time, there are no restrictions on the very implementation of HOMS.
The reason for this is related to the faster asymptotic growth of the Zeeman energy $E_Z$ with increase of $w$ as compared to the energies of anisotropy $E_A$ and scalar chiral interaction $E_K$ (see Eq.\,(\ref{E_tild}) and their discussion). Note that these conclusions are obtained in the framework of considered "linear" field model with $h(r)=r$. The physical limitations of the model itself are briefly discussed in Sec.\,\ref{sec6}.

\begin{figure}[htb!]\center
\includegraphics[width=0.48\textwidth]{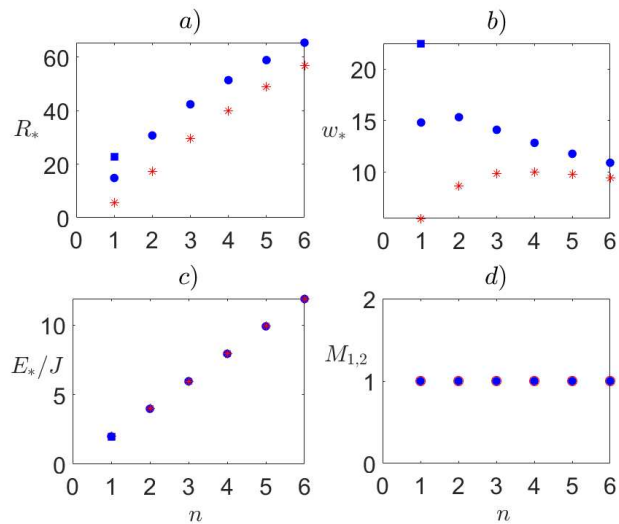} \caption{\label{4} Plots a)-c) are dependencies of the HOMS sizes and energy on the topological charge $n$. Dots represent the case $A = 0$ and $D=0$, squares -- the case $A=0$, $D=0.1$\,eV \textcolor{black}{and $n=1$}. Asterisks represent the case $A = 0.006$\,eV and $D=0$. The behavior of $M_{1}$ and $M_{2}$ (dots and circles, respectively) are the same for $A=0$ and $A \neq 0$ and shown in plot d). The other parameters in eV are $J = 160$, $K=0.076$, $H = 10^{-4}$.}
\end{figure}

\begin{figure}[htb!]\center
\includegraphics[width=0.5\textwidth]{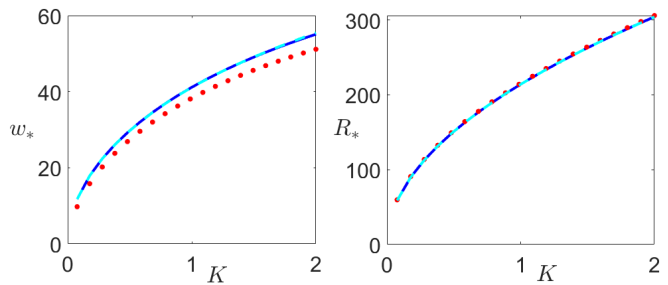} \caption{\label{5} Dependencies of the optimal HOMS sizes: the domain wall width $w_*$ and the radius $R_*$ on the amplitude of the scalar chiral interaction $K$ \textcolor{black}{at $n = 3$}. The solid and dashed curves correspond to the numerical and analytical (see (\ref{Rw_an_H})) solution of the Eqs.\,(\ref{w5_sms_ddm}), (\ref{rho_vs_w}) at $A=0$, respectively. The dots represent numerical solution of the Eqs.\,(\ref{w5_sms_ddm}), (\ref{rho_vs_w}) for $A = 0.006$\,eV. The other parameters are the same as in Fig.\,\ref{4}.}
\end{figure}

Note that the approximate solution (\ref{Rw_an_H}) of the Eq.\,(\ref{w5_sms_ddm}) has been obtained for the case $A=0$ (see Appendix D). \textcolor{black}{Wherein, in the case $A \neq 0$ the equation (\ref{w5_sms_ddm}) can be solved in the similar way, but this solution is not given here due to its complexity. On Fig.\,\ref{5} the typical dependencies of the energy and size of HOMS on the vorticity $n$ are shown. Dots represent dependencies at $A=0$, asterisks at $A \neq 0$. Also, the cases with $A=0$ and $A \neq 0$ are shown on the Fig.\,\ref{6} in $w_*$ vs. $K$ and $R_*$ vs. $K$ dependencies by the curves and dots, respectively. It can be seen that the behavior of such dependencies for $A \ll K,\,J$ and $n \geq 2$ coincide semi-quantitatively with the case $A = 0$, described analytically. Also, on the Fig.\,\ref{4}, the squares demonstrate the energy and size of the ordinary skyrmion ($n=1$), then the Dzyaloshinskii-Moriya interaction with the amplitude $D \sim K$ is taken into account. It is seen that the size of the skyrmion increases in this way without a significant change of its excitation energy.}

\section{\label{sec6} The stability of HOMS under magnetic profile variations}

Let us consider the stability of the HOMS in inhomogeneous magnetic fields beyond the linear approximation: $h(r)=r^{\beta}$, with arbitrary \textcolor{black}{$\beta \in \mathbb{R}$}. Then the functional $E$, which describes the energy of a HOMS with a sharp domain wall, has the form (see Appendix C):
\begin{eqnarray}
\label{E_gen3}
E &=& J\left(\rho + n^2\rho^{-1}\right) - D\,\rho\,w\,\delta_{n,1} + A\,\rho\,w^2 +\nonumber\\
&+&\left(E^{(\beta)}_K + E^{(\beta)}_Z\right),\\
E^{(\beta)}_Z &=& -H\left(\frac{w}{2}\right)^{\beta + 2}\Gamma_{\beta+2}\,\Li_{\beta+2}\left(-e^{2\rho},\,\textcolor{black}{\delta r}\right),\nonumber\\
E^{(\beta)}_K &=& -K\,n \,\rho^{\beta}\,w^{\beta}.\nonumber
\end{eqnarray}
Note, that in the regime $J \gg K \gg A, H$ one can give qualitative arguments concerning the existence or absence of local minima of the functional (\ref{E_gen3}). Taking into account only the term with $J$ leads to a line $\rho=n$ of degenerate local minima in the plane $\left(\,\rho\,,\,w\right)$. Since $J$ is the largest energy parameter, we can restrict the functional (\ref{E_gen3}) as $$E \to \tilde{E}(w) = E(\,w,\, \rho=n\,).$$
Then we can analyze the function of one variable, $\tilde{E}(w)$, in order to find local minima corresponding to the HOMS with topological charge $n>1$. Thus, estimating the characteristic lengths of such structures requires solving the equation:
\begin{multline}\label{E_tild}
\frac{1}{w}\frac{d\,\tilde{E}}{d\,w}= 2\,A\,n - K\,\beta\,n^{\beta + 1}\,w^{\beta-2} - \\
- H\,\frac{\beta+2}{2^{\beta+1}}\,\Gamma_{\beta+2}\,\Li_{\beta+2}\left(-e^{2n}\right)\,w^{\beta}=0,
\end{multline}
with $w \gg 1$ and $d^2E\,/\,dw^2>0$.
Analysis of the Eq.\,(\ref{E_tild}) makes it possible to find a qualitative difference between the conditions for the formation of HOMS with and without the Zeeman effects of magnetic field. \textcolor{black}{For the last case the actual solutions with $w_* \gg 1$ arise if
$$\beta < 2;~~\sign(K\,n^{\beta})=\sign(\beta\,A);~~\sign(A\,n)>0.$$} If, along with the orbital ones, we take into account the Zeeman effect of the magnetic field, there is a competition between terms on the right side of Eq.\,(\ref{E_tild}) proportional to $K$, \textcolor{black}{$A$} and $H$ (recall that $\Li_{\beta+2}\left(- e^{2\rho}\right) < 0$).

\textcolor{black}{In the case of increasing fields, $\beta > 0$, the main competition takes place between interactions with amplitudes $K$ and $H$.} So, for small $w$ we have $w^{\beta -2 } \gg w^{\beta}$ and $d\,\tilde{E}\,/\,d\,w < 0$, while for large $w$ the relations are inverse.
Thus, the right-hand side of  Eq.\,(\ref{E_tild}) changes sign from negative to positive for certain $w_*(n)$, which leads to the formation of an HOMS for arbitrary \textcolor{black}{$\beta>0$}.

\textcolor{black}{For decreased fields, $\beta < 0$, the main competition takes place between the interactions describing single-ion anisotropy ($A$) and the interactions arised from an external magnetic field ($K$ and $H$). Moreover, the details of such competition differ in cases $-2< \beta < 0$ and $\beta < -2$. If $-2 < \beta < 0$ the best way to find the HOMS solution is to fulfill the relations:
\begin{align}\label{beta_neg_cond}
\sign(A\,n)=1,~~\sign(K)=\sign(H)=-1.
\end{align}
Thus an applied magnetic field should be negative. In the case of rapidly decreased fields, $\beta < -2$, HOMS stabilization is also possible, but with the dominance of the Zeeman effects of the magnetic field.}

Thus, the conditions for the formation of HOMS are much broader if we take into account the Zeeman splitting compared to if we neglect it. However, in a uniform magnetic field ($\beta=0$), HOMS do not arise in both cases.

Moreover, one can verify that under the conditions $J\,n \gg K\,n^{\beta+1}\,w_*^{\beta-1}\,,\,H\,w_*^{ \beta+1}$ solutions of Eq.\,(\ref{E_tild}) make it possible to semi-quantitatively describe the dependence of the energy and size of the HOMS for various topological charges $n$, and the magnetic field profile degree $\beta$. For large $w_*$ and $n$ the above qualitative description is invalid. In particular, one cannot predict in this way the absence HOMS for $\beta=1$, $n > n_*$ if we neglect the Zeeman splitting.

To demonstrate the validity of the above qualitative conclusions, let us consider the question of implementation of HOMS for some particular cases:
\begin{enumerate}
    \item $\beta = 0$ -- the uniform field. In this case
\begin{eqnarray*}
\label{wang}
E^{(0)}_Z = H\left(\frac{1}{2}\rho^2 w^2 + \frac{\pi^2}{24}w^2 \right);~~E^{(0)}_K = -K\,n.
\end{eqnarray*}
Such functional describe magnetic skyrmions in the systems without chiral interactions (see  Ref.\,\cite{wang18}). So, HOMS do not arise.

    \item $\beta=1$. The skyrmions in \textcolor{black}{"linear increased"}
    field
\begin{eqnarray*}
\label{sms_ddm}
E^{(1)}_Z = H\left(\frac{1}{3}\rho^3 w^3 + \frac{\pi^2}{12}\rho w^3 \right);~~E^{(1)}_K=-K\,n \,\rho\,w.
\end{eqnarray*}
The implementation of the HOMS for such a field has been demonstrated analytically in Sec.\,\ref{sec3} and in Sec.\,\ref{sec4}.

    \item $\beta=2$. The case of a skyrmion in "quadratic" inhomogeneous magnetic field. Thus
\begin{multline*}
\label{sms_ddm_05}
E^{(2)}_Z = \frac{H}{4}\,w^4\,\left(\rho^4 + \frac{\pi^2}{2}\rho^2 + \frac{7}{15}\left(\frac{\pi}{2}\right)^4 \right);\\E^{(2)}_K=- K\,n \,\rho^2\,w^2.
\end{multline*}

    \item $\beta = 1/2$. The case of a skyrmion in "square root" inhomogeneous magnetic field. For it
\begin{multline*}
E^{(1/2)}_Z = H\,w^2\,\rho^{1/2}\,w^{1/2}\,\left(\frac{2}{5}\,\rho^2 + \frac{\pi^2}{16}\right);\\~E^{(1/2)}_K=- K\,n \,\rho^{1/2}\,w^{1/2}.
\end{multline*}\textcolor{black}
    {\item $\beta = -1$. The case of a "linear decreased" magnetic field with the functional
\begin{align*}
E_Z^{(-1)}=H\,\rho\,w\,;~~E_{K}^{(-1)}=- K\,n \,\rho^{-1}\,w^{-1}.
\end{align*}}
\end{enumerate}

The cases 2.\,--\,\textcolor{black}{5}. are visualized in Fig.\,\ref{6}.
Minima of energy functionals with $\beta =\,\,$2, $1/2$, \textcolor{black}{$-1$} were fined numerically by the gradient approach. The results of such study are shown in Fig.\,\ref{7}. It can be seen that for all considered $\beta$ there are local minima corresponding to HOMS solutions.

Moreover, it can be shown, that the HOMS states with large topological charges, $n \gg 1$, can be  have $E_*<0$ and be stable if $\beta \geq 2$ (see. Fig.\,\ref{7}).
However, it should be noted that the creation of such HOMS requires magnetic fields with a very high intensity at the skyrmion boundary, and therefore has physical limitations. Some of these limitations are briefly discussed in the next Sec.\,\ref{sec6}.

\newpage

\begin{figure}[htb!]\center\label{6}
\includegraphics[width=0.394\textwidth]{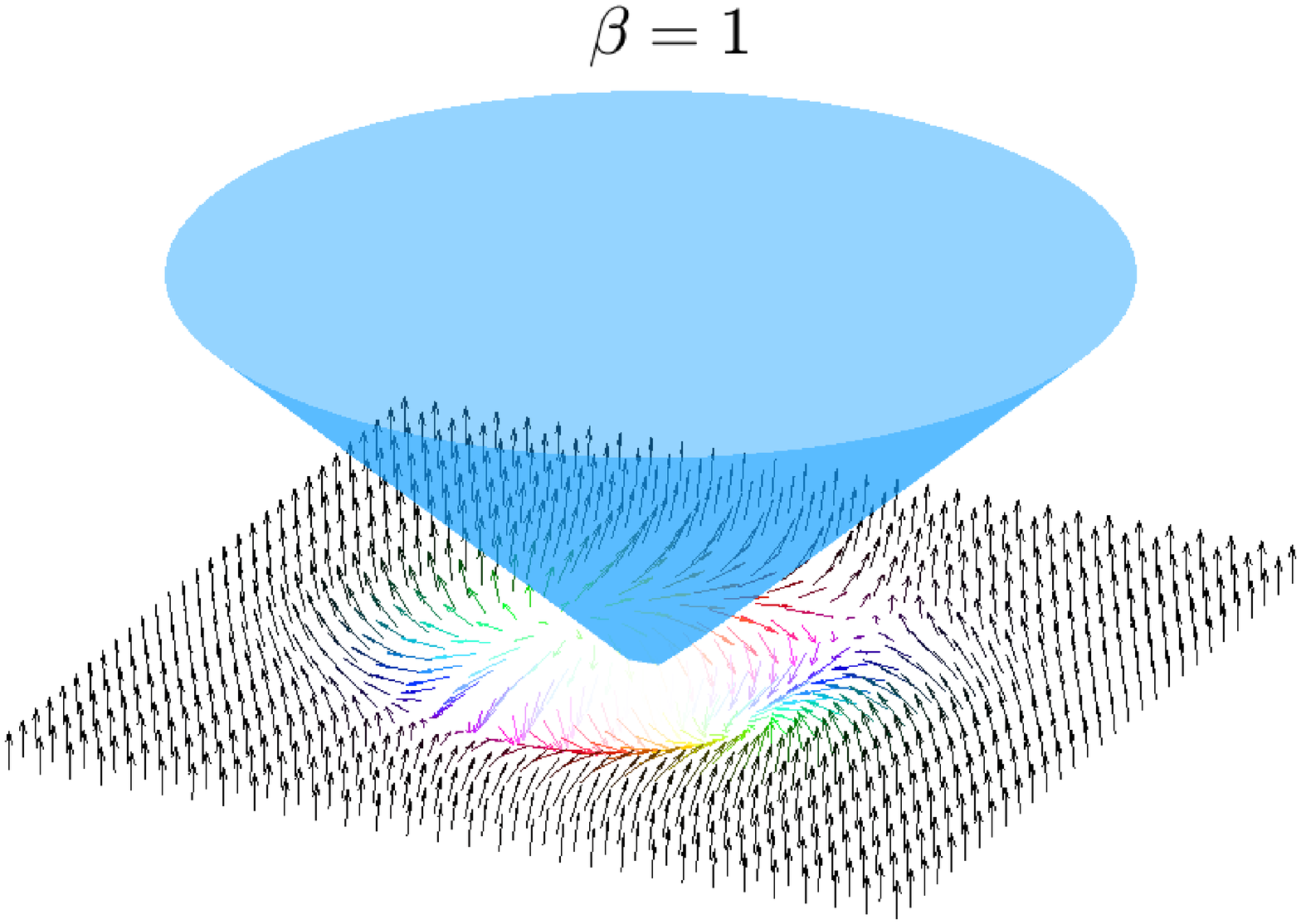}
\includegraphics[width=0.394\textwidth]{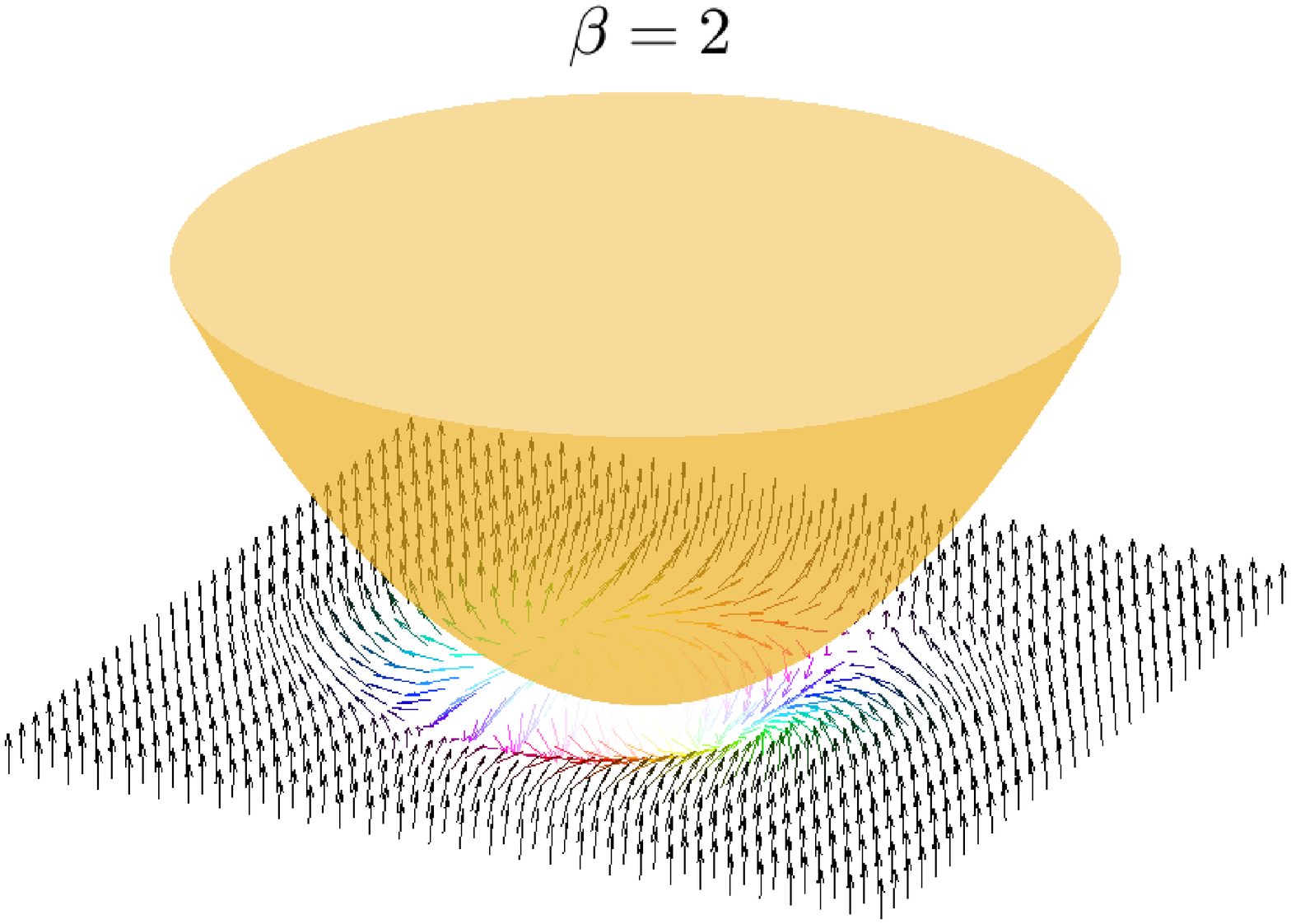}
\includegraphics[width=0.394\textwidth]{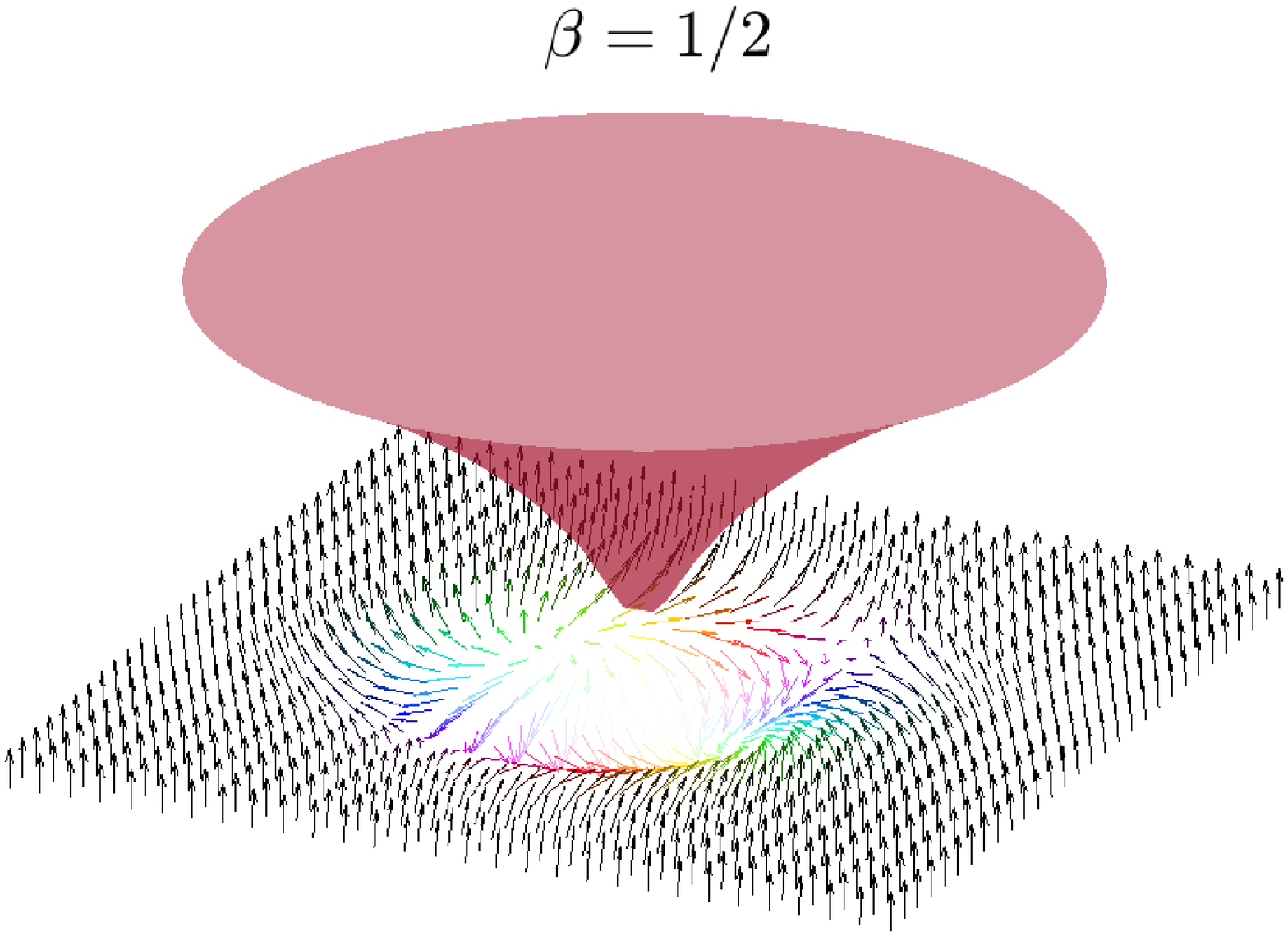}
\includegraphics[width=0.394\textwidth]{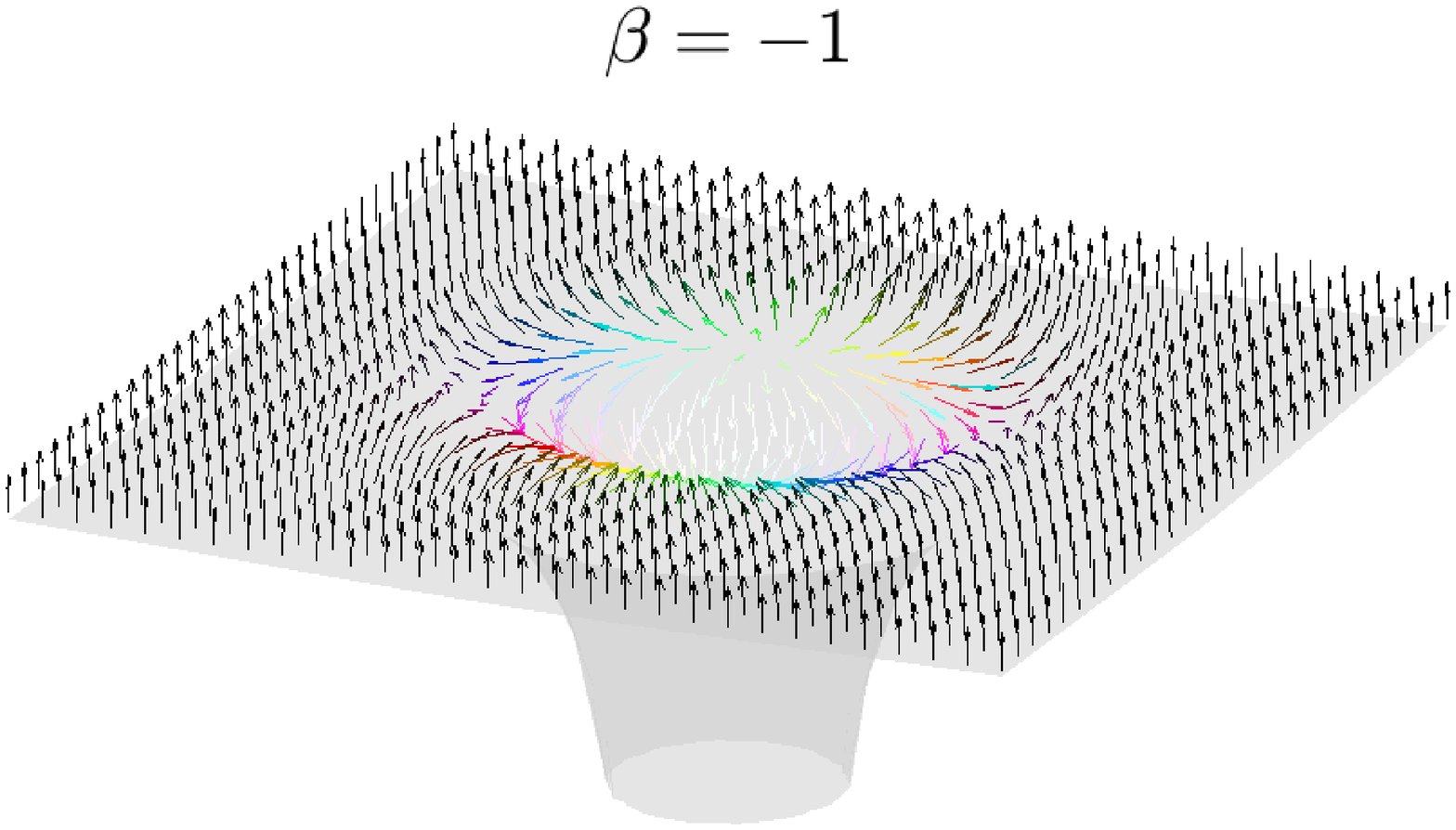}
\caption{\label{6} Visualization of the spatial profiles of HOMS in inhomogeneous \textcolor{black}{"linear increased"} ($\beta=1$), "quadratic" ($\beta=2$), "square root" ($\beta=1/2$) and \textcolor{black}{"linear decreased" ($\beta=-1$)} axially symmetric magnetic fields. The spatial profiles of the last ones are depicted in blue, orange, pink and \textcolor{black}{gray} colors, respectively. \textcolor{black}{In the latter case, the HOMS stabilization requires negative applied fields, see Eq.\,(\ref{beta_neg_cond}) and discussion below Eq.\,\ref{Rw_beta_m1}. For clarity, the modulus of the applied fields are limited in the plots.} }
\end{figure}

\begin{figure}[htb!]\center
\includegraphics[width=0.46\textwidth]{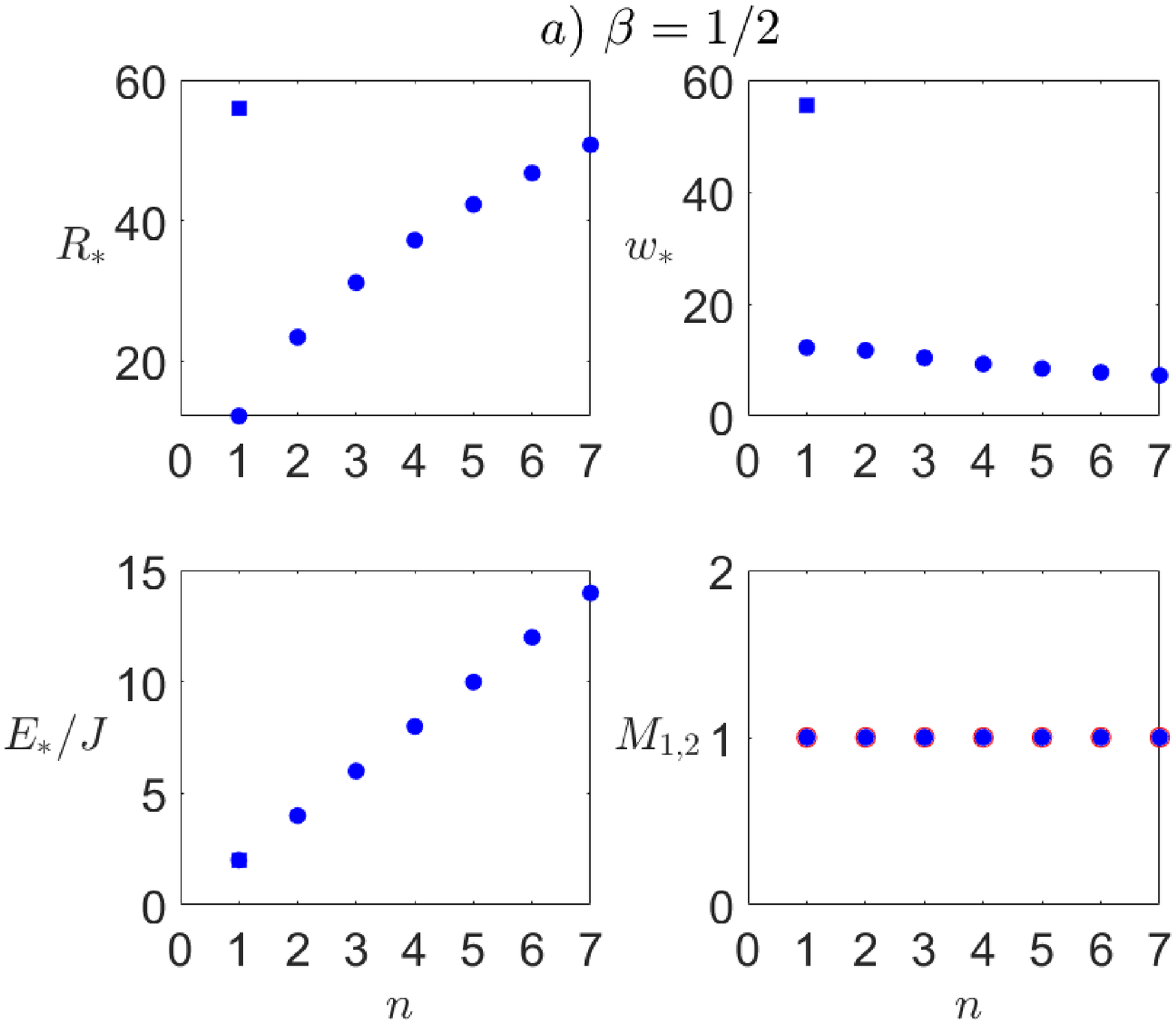}
\includegraphics[width=0.46\textwidth]{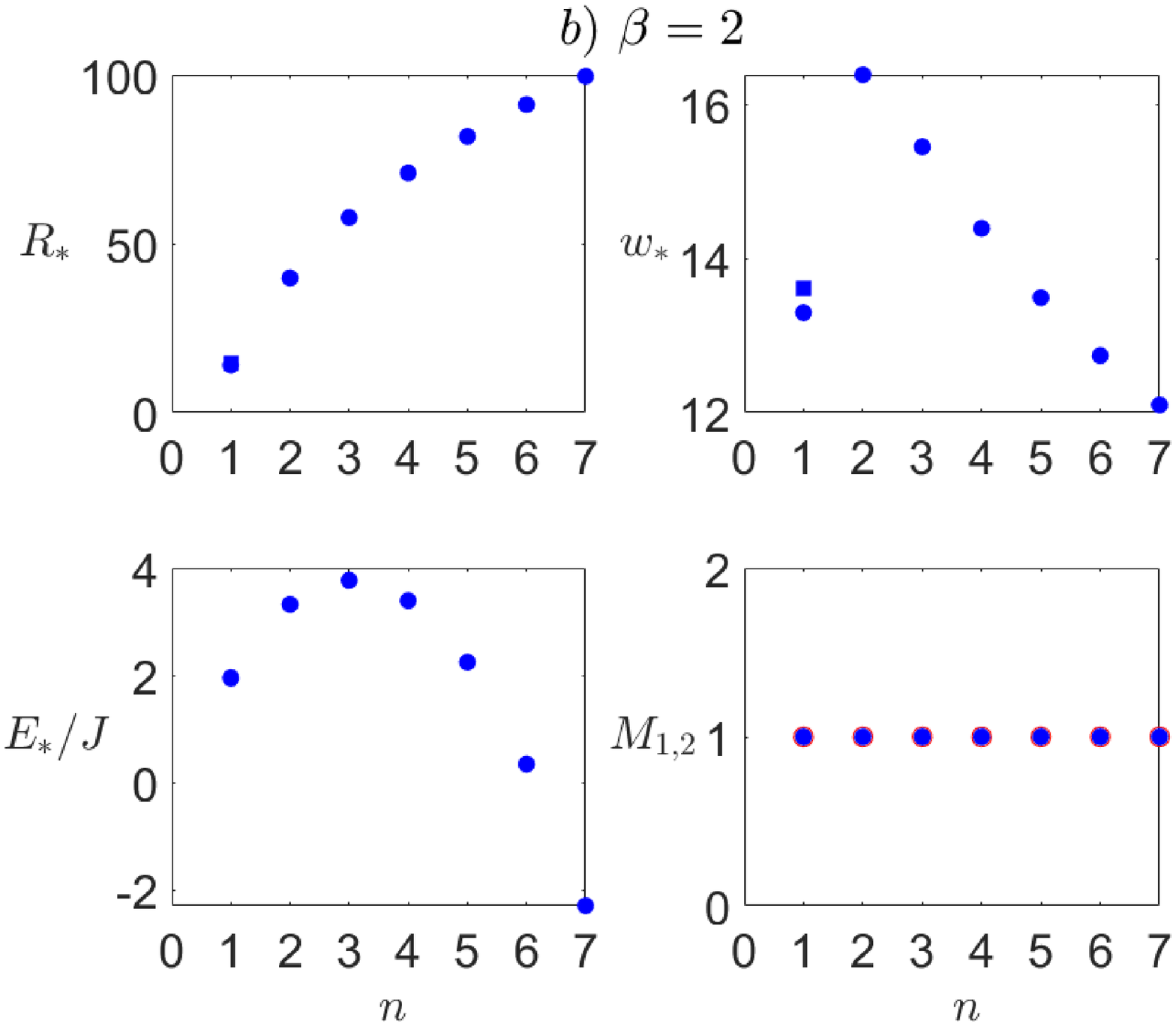}
\includegraphics[width=0.46\textwidth]{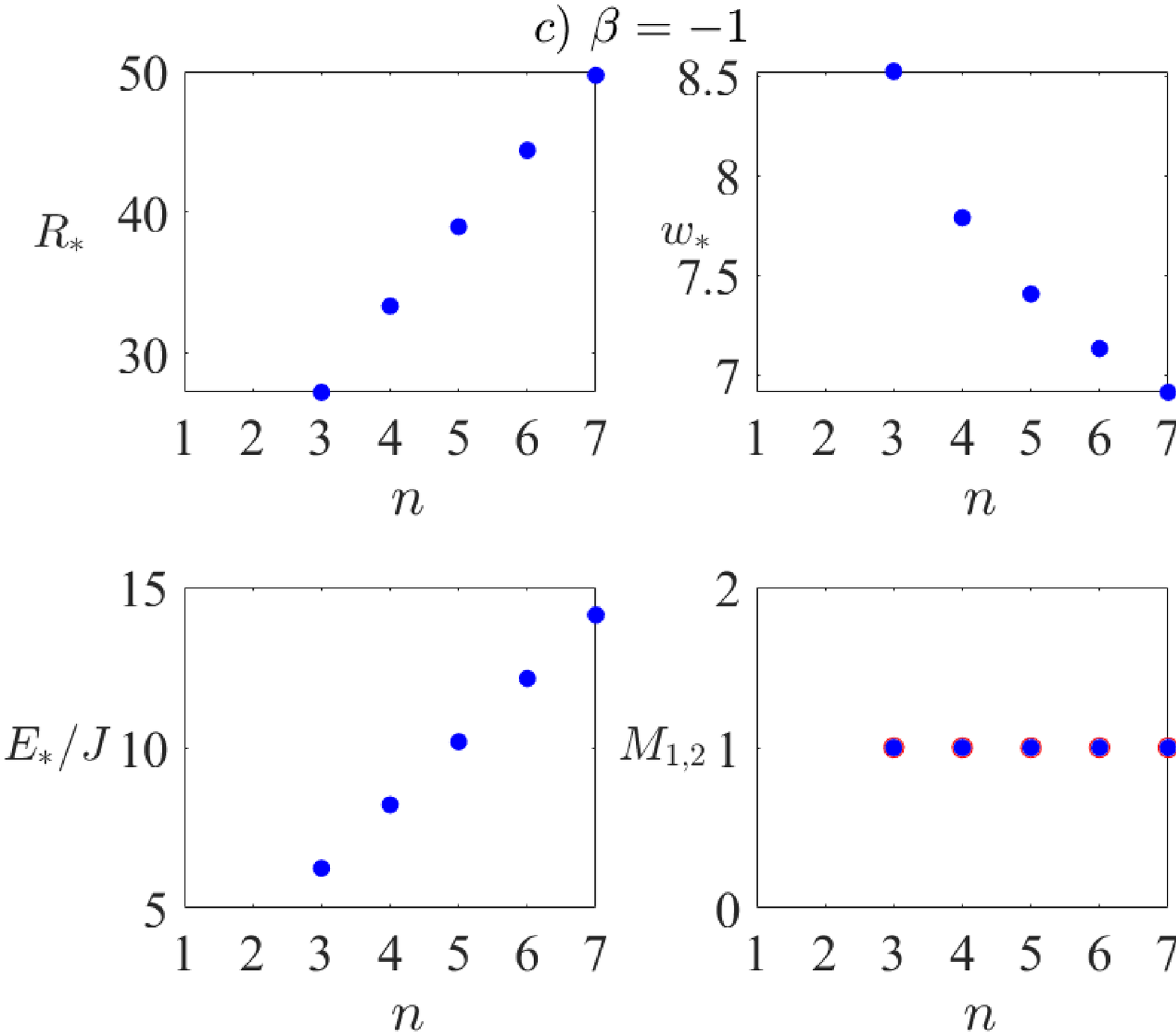}
\caption{\label{7} Dependences of the HOMS parameters on the topological charge $n$ in a) "square root" ($\beta=1/2$), b) "quadratic" ($\beta=2$) \textcolor{black}{and c) "linear decreased" non-uniform magnetic fields. In the plots a) and b) all parameters and symbols are the same as in Fig.\ref{4}. In plot c) the energy parameters in eV are: $J=17$, $K = -1$, $H=-0.012$, $A = 0.001$.}}
\end{figure}

\newpage

\textcolor{black}{Let us now consider the case of linear decreased magnetic fields, $\beta = -1$, in more detail, since the physical systems which could correspond to this case have been recently studied. It has been shown in Refs.\,\cite{dahir19, menezes19, dahir20, andriyakhina21, andriyakhina22, apostoloff22}, that an axially symmetric bound state of a magnetic skyrmion -- Pearl vortex can be realized in superconductor-chiral magnet heterostructures. If the vortex size significantly exceeds the skyrmion radius, $\lambda \gg R$, the skyrmion can be considered in the main approximation as placed in an magnetic field, ${\bf{B}} = B\,{\bf{e}}_z + B_r\,{\bf{e}}_{r}$, with linearly decreased longitudinal and radial components: $B \sim B_r \sim 1\,/\,r$ \cite{andriyakhina21, andriyakhina22, apostoloff22}. It has been shown in Ref.\,\cite{apostoloff22} that the radial field components leads to a distortion of the skyrmion texture with $n=1$, which is well described by the anzats (\ref{sk_param}) with modified skyrmion angle
\begin{eqnarray}
    \label{Theta2}
    &&{\Theta}(r) \to \tilde{\Theta}(r) = \Theta(r) + \delta\Theta_{\gamma}(r)\,\cos\Theta(r)\,,\\
    \label{dTheta}
    &&\delta\Theta_{\gamma} = \frac{\phi_0\,\mu_B\,g\,S}{8\,\pi\,\sqrt{J\,A}\,\lambda}\left[K_1\left(\frac{r}{\sqrt{J/A}}\right)-\frac{\sqrt{J/A}}{r}\right],
\end{eqnarray}
where $K_1(x)$ is the modified Bessel function of the second
kind. Let us assume that an axially symmetric structure with a skyrmion angle $\tilde{\Theta}(r)$ is realized in the case of HOMS with $n>1$. Then, assuming the system parameters corresponding to Fig.\,\ref{7}\,c) and Eq.\,(\ref{param}) it can be shown that $\delta\Theta \sim 10^{-3}$ in a wide range of variables $R$ and $w$. Thus, in our case we can use an ansatz (\ref{sk_param}) as an appropriate for our calculations: $\Theta(r) \cong \tilde{\Theta}(r)$.}

\begin{figure}[htb!]\center
\includegraphics[width=0.49\textwidth]{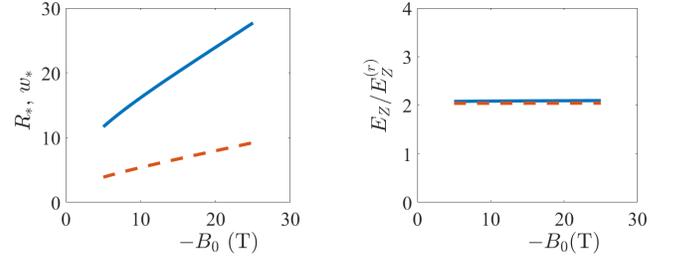} \caption{\label{9} \textcolor{black}{Left: field dependencies of radius, $R_*-$\,solid curve, and domain wall width, $w_*-$\,dashed curve, of HOMS with $n=3$. Right: the ratios of the Zeeman contributions on the interaction of HOMS with the transverse, $E_Z$, and radial, $E_Z^{(r)}$, components of the magnetic field. The solid and dashed lines correspond to the calculation of the Zeeman contributions by the Eq.\,(\ref{E_gen2}) and (\ref{EZr}) using the skyrmion angle functions, $\Theta$ and $\tilde{\Theta}$, described by the ansatz (\ref{sk_param}) and (\ref{Theta2}), respectively. In the both panels, the external magnetic field is negative; is directed opposite to the magnetization outside the HOMS boundary. The last one is marked as $-B_0(\textrm{T})$}.}
\end{figure}

\textcolor{black}{Moreover, if we take into account the contribution to the energy functional, Eq.\,(\ref{EJ2})-(\ref{EA2}) and (\ref{E_gen2}), the Zeeman term describing the interaction of magnetic system with radial magnetic field component
\begin{equation}\label{EZr}
E^{(r)}_Z = \frac{H}{2}\,\int_0^{\infty} h(r)\,\sin \Theta \,r\,dr
\end{equation}
we obtain that in a wide range of variable parameters $R$ and $w$ the ratio of the Zeeman contributions is almost constant, see the right panel of Fig.\,\ref{9}:
$$\frac{E_Z}{E^{(r)}_Z} \cong \const.$$
This ratio depends significantly only on the material parameters $J$, $D$, $K$, $A$ and $H$.
This means that in the case of $\beta = - 1$ and the considered energy hierarchy (\ref{JDKA}), description of the HOMS excitation energy by the function
\begin{equation}\label{En_beta_m1}
E^{(-1)} = J\left(\rho + n^2\rho^{-1}\right) - K\,n\, \rho^{-1}w^{-1} + A\,\rho \,w^2 +H\,\rho\, w
\end{equation}
can be considered as a suitable zero-order approximation in the problem of HOMS in the field of a Peirls vortex, even when the radial components of the field are taken into account. The radial components in this approximation lead to a renormalization of the parameter $H$ in Eq.\,(\ref{En_beta_m1}). Refinement of this approach will be the subject of further study.}

\textcolor{black}{Interestingly, in the case under consideration, the dependence of the optimal sizes of HOMS on the system parameters is also found analytically, by minimizing the function in Eq.\,(\ref{En_beta_m1}):
\begin{multline}\label{Rw_beta_m1}
R_*=\frac{w_*\,n}{\sqrt{1-A\,w_*^2\,/\,J}};~~
w_* = \frac{3\,u^{1/3}}{3\,u^{1/3} - H\,n\,/\,K + A\,/\,J};
\\
u = -\frac{A\,n}{K} + \sqrt{\left(\frac{A\,n}
{K}\right)^2+\left(\frac{H\,n}{3K}-\frac{A}{3J}\right)^3}.
\end{multline}}
\textcolor{black}{The dependencies of the HOMS sizes with $n=3$ on the induction of the external magnetic field are shown in Fig.\,\ref{9}. It can be seen that with an increase in the field strength, both the HOMS radius, $R_*$, and the domain wall width, $w_*$, increase. At the same time, the ratio $\rho_* = R_*\,/\,w_*$ also rises with an increase in both the induction amplitude, $|B_0|$, and the vorticity, $n$, which corresponding to the assumption of the analytical theory. It is important to note that the applied magnetic field should be negative, $B_0 < 0$, in order to stabilize HOMS, if the inhomogeneous fields are decreased: $\beta < 0$, see Eq.\,(\ref{beta_neg_cond}). In this case, the growing of $R_*$, $w_*$ vs. $|B_0|$ dependencies on the Fig.\,\ref{9} correspond to rise $|E_Z|$, since inside the HOMS core the magnetic field amplitude is maximum and directed along the magnetic moments. In the case of growing magnetic fields, the magnetic field strength is maximum beyond the skyrmion boundary, and to minimize the Zeeman contribution to the HOMS excitation energy, the field should be applied along the magnetic moments outside the boundary. This seems to be the main reason for the differences in the conditions of HOMS stabilization  at $\beta > 0$ and $\beta < 0$, concerning the sign of applied magnetic field.}

\textcolor{black}{Note that the numerical analysis of the functional (\ref{EF2})-(\ref{EA2}) and (\ref{E_gen2}) showed the absence of metastable HOMS states in the case $\beta = -1$ and $n=2$. Therefore, in Fig.\ref{7}\,c), the dependencies start from the values $n=3$.}

\newpage

\section{\label{sec7} Speculations about practical realizations}

As already mentioned, the HOMS stabilization models proposed above are idealized, since they consider nonuniform magnetic fields of cylindrical symmetry with a power-law dependence of their intensity on the radius, $h(r)\sim r^{\beta}$. However, even within the framework of such models, there are physical limitations associated with: HOMS morphology, the hierarchy of energy parameters, and the limiting strengths of magnetic fields. Let's consider these limitations, as well as promising physical systems in which HOMS could be observed.

Morphological limitations are related to the fact that we use the continuum description of HOMS with a sharp domain wall. Such a description assumes that the optimal sizes of structures satisfy the inequalities
\begin{eqnarray}\label{restr1}
R_{*} \gg w_{*} \gg a\,;~~2\,\pi\,R_* \gg n\,a .
\end{eqnarray}
\textcolor{black}{Restrictions on the magnetic field profiles are related to the fact that the maximum values of the applied field induction should not exceed the experimentally achievable values $B_{ex}\sim10\,$T. In the case of fields with the profiles $h(r)=r^{\beta}$, such conditions differ for the cases of increased and decreased fields
\begin{eqnarray}
\label{restr2}
\left\{ {\begin{array}{*{20}{c}}
B_0 \lesssim B_{ex},~~~~~\beta < 0,\\
B_0\,R^{\beta}_* \lesssim B_{ex},~ \beta > 0.
\end{array}} \right.
\end{eqnarray}}
The connection between $B_0$ and $H$ is defined in the Eq.\,(\ref{en_params}). Thus, the magnetic field strength must have an upper limit, at the core or the radius of the skyrmion. It can be seen from Eqs.\,(\ref{rho_vs_w})-(\ref{Rw_an_H}) that the condition
\begin{eqnarray}\label{restr3}
K\,n \sim D \gg H\end
{eqnarray}
can serve as a criterion for the constraint discussed. This expression implies that the scalar chiral interaction should have the same order of amplitude as the Dzyaloshinskii-Moriya interaction.

On the other hand, the amplitude of the scalar chiral interaction is itself proportional to the magnetic field strength in the skyrmion core: $K \varpropto H$. This condition together with Eq.\,(\ref{restr3}), gives restrictions on the parameters of the microscopic Hubbard model $U$, $t$, $\alpha$. So, the described constraints restrict the class of candidate materials for the implementation of HOMS in nonuniform magnetic fields. \textcolor{black}{Nevertheless we stress that such parameters range  still exists and corresponds to the} regime of strong electron correlations. \textcolor{black}{For instance,} it is easy to verify that for the parameters
\begin{eqnarray}\label{param}
t=0.7\,\textrm{eV};~U =5\,\textrm{eV}; ~\alpha=0.01\,\textrm{eV};\\ \nonumber
~a = 10\,\textrm{\AA};~B_{0}= 5\textcolor{black}{(22)}\,\textrm{T};~
S = 2;~g = 1,~
\end{eqnarray}
the higher-order magnetic skyrmions exist within all the considered energy functionals.
\textcolor{black}{Herein, $B_0=5\,\textrm{T}$ corresponds to the case of increasing fields, $\beta > 0$, and $B_0=22\,\textrm{T}$ describe the case of decreasing field, $\beta = -1$. The values of $t$ and $U$ in (\ref{param})  satisfy the conditions at which the effective interactions were obtained in Appendix \ref{appendix_B} and are very close to that usually used, for axample, when describing the properties of Hight-$T_c$ cuprates within the Hamiltonian of the Hubbard model \cite{Ogata_2008,VVV_DDM_MMK_AFB_2021}.}
However, the variation range of these parameters is not \textcolor{black}{too} wide if we focus on the low-energy Hamiltonian described in Appendix \ref{appendix_B}.

It can \textcolor{black}{also}  be seen from Eqs.(\ref{en_params}) and (\ref{restr3}) that the layered systems having a larger magnetic cell and a smaller Lande $g-$factor of magnetoactive ions are more preferable.
To preserve the effect of scalar chiral interactions the bandwidth of the Hubbard model should not be very small provided that the $t \ll U$ constraint is hold.

From what has been said, it follows that the implementation of stable states of the HOMS type with $n\gg 1$ and $E_*<0$ (see Fig.\,\ref{7}b)) may be problematic in practice. Indeed, the such states imply a sufficiently strong scalar chiral interaction $K$, which, in turn, requires a sufficiently strong magnetic field, $B_0$. Thus, such states would contradict the restrictions (\ref{restr1}) and (\ref{restr2}).

It is worth to stress that the conditions (\ref{restr1})-(\ref{restr3}) take place when the orbital and Zeeman effects of the magnetic field are jointly taken into account. If we consider only orbital effects of the magnetic field, then only the constraint (\ref{restr2}) remains. In this case, the limitation on magnetic field strengths is less severe, however there is a stronger limitation on the magnetic field profiles (i.e. on the degree $\beta$), as noted in the Sec.\,\ref{sec6}.

\begin{figure}[htb!]\center
\includegraphics[width=0.4\textwidth]{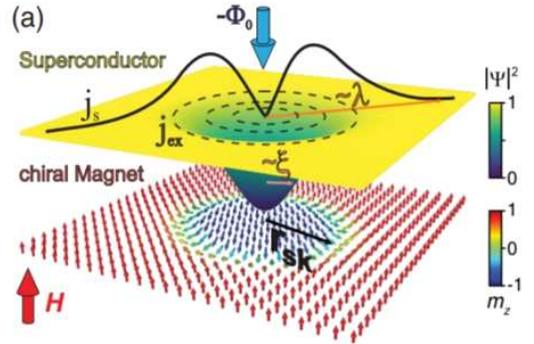} \caption{\label{8} Scheme (adopted from work \cite{petrovic21}) of a N$´$eel skyrmion with the radius $r_{sk} = R$ creating an antivortex with the flux $\Phi_0 = ch/2e$ antiparallel to the external magnetic field. The antivortex currents $j_s$
flow at radii up to $\lambda$. The superconducting order parameter $|\,\Psi\,|$ is suppressed over a length $\xi \sim R_v$ in the vortex core.}
\end{figure}

Let us briefly consider the question of the sources of an axially symmetric inhomogeneous magnetic field.
As already mentioned at the end of Sec.\,\ref{sec3}, \textcolor{black}{for the profiles with $\beta > 0$} we consider magnetic fields that do not depend on the Cartesian $z$-component. In practice, such a situation can be realized when a magnetic film is placed in a cylinder of anisotropic magnetic material. However, at present it is not clear how such a system can be used in spintronics and quantum computing devices.
In this regard, it seems \textcolor{black}{more} promising to use of a bound state of a magnetic skyrmion-superconducting vortex, \textcolor{black}{that correspond to the profile with $\beta = -1$. Recall that} such bound states can arise due to spin-orbit interaction and proximity effects \cite{hals16, baumard19}, as well as stray fields \cite{dahir19, menezes19, dahir20, andriyakhina21, andriyakhina22}. Moreover, it has been shown in recent Refs.\,\cite{andriyakhina21, andriyakhina22} that taking into account sufficiently strong stray fields can lead to the stabilization of an axially symmetric configuration of MS and superconducting vortice, see Fig.\,\ref{8}.

Practically, the bound state of magnetic and SC vortices can arise when a magnetic thin field / thin dielectric layer (for example, MnO) / and type II SC (for example, Nb) heterostructure is created. It is now claimed that such a state has been experimentally observed in the structure [Ir$_{1}$Fe$_{0.5}$Co$_{0.5}$Pt$_1$]$^{10}$\,/\,MgO\,/\,Nb \cite{petrovic21}. Thus, if a vortex state is realized in the superconductor with the vortex center at the point ${\bf{r}}_0$ and the coherence length $\xi$ (see Fig.\ref{8}), then in the spatial region $|\, {\bf{r}}-{\bf{r}}_0| \ll \xi$  the "power-law" profile of the external magnetic field $h(r)\sim a + b\,r^{-\gamma}$ is realized and the HOMS could arise. However, the solution of such a problem requires prior determination of the spatial profile of the stray magnetic field near the vortex and further application of the approach developed in this work. This problem will be the subject of further research.

In concluding this section, we note that bound states of a magnetic skyrmion - superconducting vortex are currently being studied in the context of the search for Majorana bound states localized on the MS. The reason for such a search was that the Majorana modes can be localized either on the HOMS with $n>1$ in homogeneous superconductors \cite{yang16}, or on the MS with $n=1$, but near the superconducting vortex \cite{rex19}. In the present work, we have seen that non-uniform axially symmetric magnetic field
by itself can be a source of HOMS with $n \geq 2$. This case is close to the field profile in the SC vortex. Since there exist today technologies for controlling the position of both superconducting vortices and skyrmions, the another possible application of HOMS seems to be the creation and manipulation of the Majorana modes, their braiding and quantum computations \cite{zlotnikov21, gungordu22}.

\section{\label{sec8} Summary}

The \textcolor{black}{mechanism} of stabilizing higher-order magnetic skyrmions (HOMS) in 2D systems \textcolor{black}{due to orbital effects of the non-uniform magnetic field} is demonstrated. To estimate the characteristic energies of orbital contributions, we derive effective spin-spin interactions in the framework of the 2D Hubbard model in an external magnetic field and with account for the Rashba spin-orbit coupling. It turned out that in the regime of strong electron correlations and relatively weak spin-orbit coupling, effective spin interactions include scalar and vector chiral contributions, which can be of the same order of magnitude. The vector chiral interaction, or the Dzyaloshinskii-Moriya interaction, is determined by the spin-orbit coupling constant and depends on the material characteristics. The contribution of the scalar chiral interaction depends on the spatial profile of the external magnetic field. For a homogeneous fields, it just shifts the reference energy by a value proportional to the topological charge of the spin configuration. In the case of non-uniform field the scalar and vector chiral terms can either compete or complement each other.

We consider in details the question of the competition of chiral interactions for the case of an axially symmetric "linear" magnetic field in the context of the formation and stabilization of HOMS. We assume the field strength textcolor{black}{takes a minimum (maximum) value} at the center of the skyrmion and linearly increases \textcolor{black}{(decreases)} with distance from its center. It turned out that for magnetic skyrmions with topological charge $|Q|=1$, scalar and vector chiral interactions and either complement \textcolor{black}{or compete} each other. More interesting was the problem of stabilization of axially symmetric higher-order skyrmions with $Q>1$. It turned out that their realization in an inhomogeneous field is possible, \textcolor{black}{even if only orbital field effects are taken into account}. Moreover, the dependence of the characteristic parameters of HOMS on the topological charge or the strength of the applied magnetic field is nontrivial. To describe it we have developed analytical theory of HOMS for "linear" magnetic field.

Further, within adopted
hierarchy of effective parameters, we drew conclusions
that qualitatively generalize the results for "linear in $r$" magnetic field to the case of power degree fields, $B(r) \sim r^{\beta}$ with \textcolor{black}{$\beta \in \mathbb{R}$}. Specifically, it is concluded that consideration of the Zeeman effect of the magnetic field in addition to the orbital one significantly expands the possibilities of HOMS stabilization. Moreover, it is shown that
with \textcolor{black}{$\beta>1$}
the HOMS could be both metastable and stable.

The found features can serve as a platform for the theoretical and experimental search for higher-order magnetic skyrmions in layered strongly correlated materials by placing them in inhomogeneous magnetic field.
In this respect a promising system
is the bound state of a magnetic skyrmion--superconducting vortex. Recently, such states have been experimentally discovered in hybrid structures [Ir$_{1}$Fe$_{0.5}$Co$_{0.5}$Pt$_1$]$^{10}$\,/\,MgO\,/\,Nb \cite{petrovic21}. Moreover, recent theoretical works have shown that the stray magnetic fields induced by Pearl vortices can serve as a source of coaxial configuration of superconducting and magnetic vortices on top of each other in hybrid structure: superconductor\,/\,thin dielectric layer\,/\,chiral ferromagnet. With this in mind, as well as the results of the present work, it seems interesting to consider stabilization of the HOMS in the stray fields of a superconducting vortex, taking into account the orbital and Zeeman effects of the magnetic field.

The higher-order magnetic skyrmions themselves may be of practical interest for searching for Majorana bound states \cite{yang16} on them. So, it is known that the Majorana modes can be localized either on the HOMS with $n>1$ in homogeneous superconductors, or on the MS with $n=1$, but near the superconducting vortex. In the present work, we have seen that a superconducting vortex by itself can serve as a source of HOMS with $n \geq 2$. In the latter case, HOMS, which could be carriers of Majorana modes, may have high prospects for creation of quantum computing devices, since such skyrmions can be quite simply moved in two dimensions, realizing the braiding operations.

In conclusion, we note that in a recent paper \cite{kipp21} a close problem on the influence of the scalar chiral interaction on the features of the spin Hall effect was considered. The spin Hall effect is usually due to the existence of magnetic skyrmions in the system. However, the theoretical description of Ref.\,\cite{kipp21} was carried out at the phenomenological level without analyzing the details of chiral magnetic structures, such as magnetic skyrmions.

\begin{acknowledgments}
The authors thank prof. I. S. Burmistrov, S. Apostoloff, E. S. Andriyakhina, M.N. Potkina and A. D. Fedoseev for fruitfull discussions. S.M.S. thanks for the support the Foundation for the Advancement of Theoretical Physics and Mathematics “BASIS” (Grant No. 20-1-4-25-1) and Council of the President of the Russian Federation for Support of Young Scientists and Leading Scientific Schools (Grant No. MK-4687.2022.1). The analytical solution of systems of nonlinear algebraic equations was carried out by S.V.A. with the support by Krasnoyarsk Mathematical Center and financed by the Ministry of Science and Higher Education of the Russian Federation (Agreement No. 075-02-2023-936).

\end{acknowledgments}

%\newpage

\appendix

\section{Perturbation theory for degenerate spectrum }\label{appendix_A}

Let us write down the original Hamiltonian $\mathcal{H}$, as a sum of terms of zero, first and second order of smallness, and denote corresponding terms by the lower indices $0$, $1$ and $2$:
\begin{eqnarray}\label{H_tot}
\mathcal{H} = \mathcal{H}_{0} + \mathcal{V}_1 + \mathcal{V}_2.
\end{eqnarray}
Here $\mathcal{H}_{0}$ is unperturbed Hamiltonian and $\mathcal{V}_j$ $(j=1,2)$ -- operators describing weak interactions.

As a basis in the Hilbert space of the operator $\mathcal{H}$ it is convinient to choose manybody eigenstates $|\,m\,\rangle$ of the Hamiltonian $\mathcal{H}_{0}$: $\mathcal{H}_{0}|\,m\,\rangle=E_m|\,m\,\rangle$.
An important assumption for the development of the perturbation theory is the existence of a large energy gap in the spectrum of eigenvalues $E_m$.
The subspace of states with eigenvalues bellow the gap (the, so called, "low-energy"\ sector of the Hilbert space) will be denoted further as $\mathcal{M}$ and corresponding eigenstates and eigenvalues will be numbered by the symbol $m$.
For numbering states with eigenenergies above the gap (the "high-energy"\ sector)  we will use the symbol $l$, and the  subspace of such states will be denoted as $\mathcal{L}$.
Note that both the states $|m\rangle \in \mathcal{M}$ and the states  $|l\rangle \in \mathcal{L}$ can be degenerate but not all the eigenvalues $E_m$ (and as well $E_l$) must necessarily be equal to each other.

Using the manybody states $|m\rangle$ we can define projection operator $P$ onto the low-energy sector $\mathcal{M}$ as:
$P = \sum_{m \in \mathcal{M}} X^{mm}$,
with $X^{mm}=|m\rangle\langle m|$ being the Hubbard operators.
The projection operator allows to divide interactions $\mathcal{V}_j$ ($j=1,2$) in the Hamiltonian (\ref{H_tot}) into two parts: $\mathcal{V}_j=\mathcal{\bar{V}}_j+\mathcal{\bar{\bar{V}}}_j$. The first part ${\mathcal{\bar{V}}_j}$, consisting of two terms
\begin{eqnarray}\label{H1}
\mathcal{\bar{V}}_j = P\,\mathcal{V}_j\,P+(1-P)\,\mathcal{V}_j\,(1-P),
\nonumber\\
P\,\mathcal{V}_j\,P = \sum_{m,m'\in \mathcal{M}}\left(\mathcal{V}_j\right)_{m,m'}\hphantom{l}X^{mm'},\nonumber\\
(1-P)\,\mathcal{V}_j\,(1-P)=\sum_{l,l'\in \mathcal{L}}\left(\mathcal{V}_j\right)_{l,l'}\hphantom{l}X^{ll'},
\end{eqnarray}
does not mix the low- and high-energy sectors of the Hilbert space and hence is called the diagonal part.
The second part $\mathcal{\bar{\bar{V}}}_j$, also consisting of two terms
\begin{eqnarray}\label{H2}
\mathcal{\bar{\bar{V}}}_j = (1-P)\,\mathcal{V}_j\,P+P\,\mathcal{V}_j\,(1-P),
\nonumber\\
(1-P)\,\mathcal{V}_j\,P = \sum_{\substack{m \in \mathcal{M} \\ l \in \mathcal{L}}}
\left(\mathcal{V}_j\right)_{l,m}\hphantom{l}X^{lm},
\nonumber\\
P\,\mathcal{V}_j\,(1-P) = \sum_{\substack{m \in \mathcal{M} \\ l \in \mathcal{L}}}
\left(\mathcal{V}_j\right)_{m,l}\hphantom{l}X^{ml},
\end{eqnarray}
is non-diagonal, since it does mix the sectors $\mathcal{M}$ and $\mathcal{L}$. In equations (\ref{H1}) and (\ref{H2}) the matrix elements $\langle m|\mathcal{V}_j |l \rangle$ of the operators $\mathcal{V}_j$ are denoted as $\left(\mathcal{V}_j\right)_{m,l}$.

Consider the following unitary transformation of the Hamiltonian $\mathcal{H}$
\begin{eqnarray}\label{U_trans}
\mathcal{H}\to \tilde{\mathcal{H}}=e^{-S}\,\mathcal{H}\,e^{S}&=&\mathcal{H} + [\,\mathcal{H}, S\,] + \frac{1}{2}\big[\,[\,\mathcal{H},\,S\,],\,S\,\big] +\nonumber\\ &+&\frac{1}{6}\Big[\,\big[\,[\,\mathcal{H},\,S\,],\,S\,\big],S\,\Big]+\ldots
\end{eqnarray}
We will assume that the operator $S$ in the formular (\ref{U_trans}) is non-dioganal
\begin{eqnarray}\label{S_tot_X}
S &=& \sum_{\substack{m \in \mathcal{M} \\ l \in \mathcal{L}}}
\left[\left(S\right)_{m,l}\hphantom{l}X^{ml} + \left(S\right)_{l,m}\hphantom{l}X^{lm} \right],
\end{eqnarray}
and its decomposition starts with terms of the first order of smallness:
\begin{eqnarray}\label{S_tot}
S &=& S_1+S_2+S_3 + \ldots.
\end{eqnarray}

Substitute expresion (\ref{H_tot}) and (\ref{S_tot}) into the series (\ref{U_trans}) and retain only those terms, whose order of smallness is not higher than three.
In the, obtained in this way, expression for $\tilde{\mathcal{H}}$ we want to get rid of non-diagonal terms by imposing the following conditions on the operators $S_1$ and $S_2$:
\begin{eqnarray}\label{S1_eq}
\mathcal{\bar{\bar{V}}}_1 + [\, \mathcal{H}_{0}\, , \, S_1\,] = 0,
\\ \label{S2_eq}
\mathcal{\bar{\bar{V}}}_2 + [\,\mathcal{H}_{0}\,,\, S_2\,] + [\, \mathcal{\bar{V}}_1   ,\, S_1\,]= 0.
\end{eqnarray}
As a result, in the Hamiltonian $\tilde{\mathcal{H}}$ up to the third order only the diagonal terms remain.
Projecting out, at last, the high-energy processes we are left with operators acting only within the low-energy sector $\mathcal{M}$ of the Hilbert space and thus forming the required effective Hamiltonian
\begin{eqnarray}\label{H_eff}
\mathcal{H}_{eff} =P\,\mathcal{H}\,P+ ~~~~~~~~~~~  \nonumber\\
 +\frac{1}{2}\,P\left( \big[ \,\mathcal{\bar{\bar{V}}}_1\, , \, S_1 + S_2\, \big] +  \big[ \,\mathcal{\bar{\bar{V}}}_2\, , \, S_1 \big] \right)P.
\end{eqnarray}

From the operator equation (\ref{S1_eq}) follow the equations for the matrix elements of the operator $S_1$:
\begin{eqnarray}\label{S1_eq_matr}
\left(S_1\right)_{m,l} = \frac{\left(\mathcal{V}_1\right)_{m,l}}{E_{l} - E_{m}}\,,
~
\left(S_1\right)_{l,m} = -\frac{\left(\mathcal{V}_1\right)_{l,m}}{E_{l} - E_{m}}\,.
\end{eqnarray}
Here we took advantage of the equalities
$\mathcal{H}_{0}|\, m \, \rangle = E_{m}|\, m \, \rangle$ and
$\mathcal{H}_{0}|\, l \, \rangle = E_{l}|\, l \, \rangle$.

Let's represent the equation (\ref{S1_eq_matr}) again in the operator form by introducing operators $\mathcal{O}$ and $\mathcal{\bar{O}}$, acting on the Hubbard operators as follows:
\begin{eqnarray}\label{Q_symbs}
\mathcal{O} \, X^{lm} = \frac{X^{lm}}{E_{l} - E_{m}}\,;
\phantom{aa}
X^{ml} \mathcal{\bar{O}}  = \frac{X^{ml}}{E_{l} - E_{m}}.
\end{eqnarray}
It can be easily verified that both these operators may be represented as $\left(\mathcal{H}_{0} - K\mathcal{H}_{0}K\right)^{-1}$, where $K$ is the Hermitian conjugation operator.
The difference in notations of the operators $\mathcal{O}$ and $\bar{\mathcal{O}}$ is to emphasize that their action is directed to the right and left, respectively.
Taking into account the definitions (\ref{Q_symbs}), from the expression (\ref{S1_eq_matr}) one can written:
\begin{eqnarray}\label{S1_eq_op}
S_1 = -\mathcal{O}\,(1-P)\,\mathcal{V}_1\,P + P\,\mathcal{V}_1\,(1-P)\,\mathcal{\bar{O}}.
\end{eqnarray}
Note also that the property $\left(\mathcal{O}\,\mathcal{H}_{2}\right)^{+}=\mathcal{H}_{2}\,\mathcal{\bar{O}}$
 implies the anti-hermitiancy of  $S_1$, as it should be.

 Similarly, from the condition (\ref{S2_eq}) for $S_2$, and definitions (\ref{Q_symbs}) for operators $\mathcal{O}$ and $\bar{\mathcal{O}}$, it is easy to get:
\begin{eqnarray}\label{S2_eq_op}
S_2 &=& \mathcal{O}\left((1-P)\mathcal{V}_1 (\mathcal{O}((1-P)\mathcal{V}_1 P))\right) -
\nonumber\\
&~& -\mathcal{O}\left((\mathcal{O}((1-P)\mathcal{V}_1 P))\mathcal{V}_1 P\right)-
\nonumber\\
&~& - \mathcal{O}((1-P)\,\mathcal{V}_2P)
-h.c.~.
\end{eqnarray}
Substituting the expressions (\ref{S1_eq_op}) and (\ref{S2_eq_op}) into formula (\ref{H_eff}), we obtain the final expression for the effective Hamiltonian acting in the low-energy subspace $\mathcal{M}$:
\begin{multline}
\mathcal{H}_{eff} = P\mathcal{H}P
-\frac{1}{2}P\bigg[
\mathcal{V}_1(1-P) \mathcal{O} (1-P) \mathcal{V}_1 +
\\
+\mathcal{V}_1(1-P) \mathcal{O} (1-P) \mathcal{V}_2
+\mathcal{V}_2(1-P) \mathcal{O} (1-P) \mathcal{V}_1+ h.c.\bigg]P
+\\
+\frac12 P\mathcal{V}_1 \mathcal{O} \bigg[ (1-P)\mathcal{V}_1
\bigg(\mathcal{O}\big((1-P)\mathcal{V}_1P\big)\bigg)-
\\
-\bigg( \mathcal{O}\big((1-P)\mathcal{V}_1P\big) \bigg)
\mathcal{V}_1P\bigg]+ h.c.~
\end{multline}\label{H_eff_fin}

Consider the expression (\ref{H_eff_fin}) in the case of a degenerate (in the absence of perturbation) lower sector $\mathcal{M}$ with energy $E_{0}$.
In this case, the structure of the effective Hamiltonian (\ref{H_eff_fin}) makes it possible to simplify the representation for the operators $\mathcal{O}$ and $\mathcal{\bar{O}}$  by writing them both as $\left(\mathcal{H}_{0} - E_0 \right)^{-1}$ what allows to omit any prescriptions about their action direction.
Moreover, assuming that in the first order of perturbation theory the degenerate level $E_0$ remains degenerate, shifting by the value $\mathcal{E}_1$:
\begin{align*}\label{H_eff_first_ord}
\mathcal{E}_1= P\mathcal{V}_1P,
\end{align*}
it is easy to show that the effective Hamiltonian up to the third order is written as
\begin{multline*}
\mathcal{H}_{eff} = P\bigg[\mathcal{H} -\mathcal{V}_1 (1-P)\mathcal{O}(1-P)\mathcal{V}_1-
\\
-\mathcal{V}_1 (1-P)\mathcal{O}(1-P)\mathcal{V}_2
-\mathcal{V}_2 (1-P)\mathcal{O}(1-P)\mathcal{V}_1 + \\
+\mathcal{V}_1 (1-P)\mathcal{O}(\mathcal{V}_1 - P\mathcal{V}_1 - \mathcal{E}_1)\mathcal{O}
(1-P)\mathcal{V}_1\bigg]P.
\end{multline*}
This expression was obtained in \cite{tyablikov65, bogoliubov12} within a different formalism, which is correct only if the ground state of the system is degenerate.

\section{Effective spin interactions in the 2D Hubbard model with spin-orbit coupling}\label{appendix_B}

We apply the general formalism developed above to study effective interactions in the 2D Hubbard model with a Rashba SOI in the external magnetic field. Hamiltonian of this model we write down in the form:
$\mathcal{H}=\mathcal{H}_0+\mathcal{V}_1$, where
\begin{eqnarray}
\mathcal{H}_0 &=& \sum_{f} \psi_f^+((\varepsilon_0-\mu)\tau^0-h\tau^z)\psi_f+\mathcal{H}_U,
\\ \label{V_SOI}
\mathcal{V}_1 &=& \sum_{\langle f,g\rangle} \psi_f^+\hat{t}_{fg}\psi_g;~~\hat{t}_{fg}=t_{fg} \tau^0 + i \alpha \varepsilon_{\mu\nu z}\tau^{\mu}\,d^{\nu}_{fg}.~~~~
\end{eqnarray}
Here the two-component operator $\psi_f$
is defined as:
$\psi_f = \left(c_{f\uparrow},c_{f\downarrow} \right)^T$, where $c_{f\sigma}(c^+_{f\sigma})$ is the annihilation (creation) operator of an electron on the site $f$ with spin projection $\sigma~(=\pm 1/2)$, $\varepsilon_0$ -- bare on-site energy, $\mu$ -- chemical potential, $h=\mu_B H$ with $\mu_B$ being Bohr magneton and $H$ -- magnetic field. Pauli matrices are defined in the usual way
\begin{eqnarray}
\tau^x =
\left(\begin{array}{cc} 0 & 1 \\ 1 & 0 \end{array}
\right),~
\tau^y=
\left(\begin{array}{cc} 0 & -i \\ i & 0 \end{array}
\right),~
\tau^z=
\left(\begin{array}{cc} 1 & 0 \\ 0 & -1 \end{array}
\right),~~~~
\end{eqnarray}
and $\tau^0$ -- the unit matrix. Tunneling integral $\hat t_{fg}$ is supposed to be zero only for nearest sites $f$ and $g$. Therefore, summation in the formula (\ref{V_SOI}) is constrained only by the nearest neighbors, as indicated by the angle brackets. Diagonal part $t_{fg}$ of the matrix tunneling integral $\hat t_{fg}$ describes direct electron hoppings, and the non-diagonal part --- hoppings due to the spin-orbit interaction with $\alpha$ being intesity of SOI. In the definition of $\hat t_{fg}$ the symbol $\varepsilon_{\mu\nu z}$ ($\mu,\nu=\{x,y,z\}$) denotes antisymmetric Levi-Civita tensor and ${\bf d}_{fg}$ -- unit vector, connecting the sites $f$ and $g$. The operator $\mathcal{H}_U$ takes into account the local Coulomb interaction (with energy $U$) of two electrons with opposite spin projections.

Consider the system at half filling and in the regime of strong electron correlations $U\gg t,\,\alpha,\,h$.
As a "low-energy" sector $\mathcal{M}$, we will consider the space of homeopolar states with one electron at each site. The "high-energy" sector  $\mathcal{L}$ will be formed out of states for which at least one site have two or no electrons.
The projection operator on the sector $\mathcal{M}$ can be written in the form:
\begin{eqnarray}\label{P}
P = \prod_{f}\sum_{\sigma}X_{f}^{\sigma\sigma}.
\end{eqnarray}
Here Hubbard operators are defined as
$X_{f}^{nm}=|f,n\rangle\langle f,m|$  \cite{hubbard65, zaitcev75, zaitcev04, valkov01} and describe transitions from the manybody state $|f,m\rangle$ to the state $|f,n\rangle$ on the site $f$ with quantum numbers $m$ and $n$ respectively.
Hubbard operators $X^{mn}_{f}$ and $X^{pq}_{g}$ on different sites anticommute if in the result of both  transitions $|n\rangle\to |m\rangle$ and $|p\rangle\to|q\rangle$ the change in the number of fermions is odd, otherwise they commute.

An important difference between the operators $X^{nm}$, introduced in the Appendix A, and the operators $X^{nm}_f$ should be noted. While operators $X^{nm}$ are defined using many-body states of the entire system, operators $X^{nm}_f$ are constructed with many-body states related to only one site $f$.

In our case the basis of states on the site $f$ includes four states: the state $|f,0\rangle$ with no electrons, two states $|f,\sigma\rangle$ describing one electron with spin $\sigma$ and the state $|f,2\rangle$ with two electrons with opposite spin projections.

The electron annihilation operator on the site $f$ with spin projection $\sigma$ can be expressed in terms of Hubbard operators:
\begin{equation}\label{Fermi2Hubbard}
c_{f\sigma}=X_{f}^{0\sigma}+2\sigma X^{\bar{\sigma}2}_f.
\end{equation}
In this representation the operator of Coulomb repulsion$\mathcal{H}_U$ takes a particularly simple form:
\begin{equation}
    \mathcal{H}_U=U\sum_f X^{22}_f.
\end{equation}

According to the general formular (\ref{H_eff_fin}) we can write:
\begin{equation}
    \mathcal{H}_{eff} = P\mathcal{H}P + \delta\mathcal{H}_{2} + \delta\mathcal{H}_{3},
\end{equation}
where operators of the effective interactions $\delta\mathcal{H}_{2}$ and $\delta\mathcal{H}_{3}$, in the 2nd and 3rd orders of perturbation theory accordingly, have the form:
\begin{eqnarray}\label{H_eff_gen}
\delta\mathcal{H}_{2} &=& - \frac{1}{2}P\mathcal{V}_1(1-P) \mathcal{O}(1-P) \mathcal{V}_1P + h.c., \\
\delta\mathcal{H}_{3} &=& \frac12 P\mathcal{V}_1 \mathcal{O} \bigg[ (1-P)\mathcal{V}_1
\bigg(\mathcal{O}\big((1-P)\mathcal{V}_1P\big)\bigg)\bigg]+h.c.~.
\nonumber
\end{eqnarray}
Note that the second term in the square brackets of the last line in equation (\ref{H_eff_fin}) identically equals to zero due to the homeopolarity condition.

Taking into account the expressions (\ref{P}), (\ref{Fermi2Hubbard}) and representing the operator $\mathcal{O}$ as $\left(\mathcal{H}_{0} - K\mathcal{H}_{0}K\right)^{-1}$, we come to the following expression for the effective interaction $\delta\mathcal{H}_{2}$:
\begin{eqnarray}
    \label{H2_spinor}
&~&\delta\mathcal{H}_{2} = -\frac{1}{2}\sum_{\langle\,f,g\,\rangle}Sp\Big[\,\hat{X}_f\,\hat{t}_{fg}\,\hat{Y}_g\,\tilde{t}_{gf}\,\Big]+h.c.~,
\end{eqnarray}
written in terms of the matrices:
\begin{eqnarray}\label{XY_matrices}
\hat{X}_f = \left( \begin{array}{*{20}{c}}
X_f^{\uparrow\uparrow} & X_f^{\downarrow\uparrow}  \\
X_f^{\uparrow\downarrow} & X_f^{\downarrow\downarrow}  \\
\end{array} \right) ,~
\hat{Y}_g = \left( \begin{array}{*{20}{c}}
X_g^{\downarrow\downarrow} & -X_g^{\downarrow\uparrow}  \\
-X_g^{\uparrow\downarrow}  & X_g^{\uparrow\uparrow}   \\
\end{array} \right),~~~~~\\
\tilde{t}_{gf} = \left( \begin{array}{*{20}{c}}
(\hat t_{gf})_{11}\,/\,U & (\hat t_{gf})_{12}\,/\,\left(U-2h\right)  \\
(\hat t_{gf})_{21}\,/\,\left(U+2h\right) & (\hat t_{gf})_{22}/U   \\
\end{array} \right).~~~~~ \nonumber
\end{eqnarray}
These matrices can be expressed via Pauli matrices:
\begin{eqnarray}\label{XY_S}
\hat{X}_f = \frac{\tau^0}{2}+ {\bf{S}}_f {\bf{\tau}},~~
\hat{Y}_g = \frac{\tau^0}{2}-{\bf{S}}_g\,{\bf{\tau}},
\end{eqnarray}
$$\tilde{t}_{gf} = \frac{1}{U}\bigg[t_{gf}\tau_0 + \frac{i\,\alpha}{1-x^2}\varepsilon_{\mu\nu z}\,\tau^{\mu}\,d^{\nu}_{gf}-\frac{\alpha\,x}{1-x^2}\left({\bf{d}}_{gf}\cdot \bf{\tau}\right) \bigg],$$
where $x = 2h/U$ and ${\bf S}_f$ is the $1/2$-spin operator on the site $f$. In Eq.\,(\ref{XY_S}) homeopolarity condition was used.

Using (\ref{XY_matrices}) and  (\ref{XY_S}) in Eq.\,(\ref{H2_spinor}) we obtain expression for  $\delta\mathcal{H}_{2}$ for
effective spin interactions in the Hubbard model in the second order of perturbation theory:
\begin{multline}\label{Heff2_tot}
\delta\mathcal{H}_{2} = \sum_{\langle fg\rangle}\bigg[ -\bigg(\frac{t^2}{2U}+\frac{\alpha^2}{2U(1-x^2)}\bigg) + \\
+ \bigg(\frac{2t^2}{U}-\frac{2\alpha^2}{U(1-x^2)}\bigg){\bf{S}}_f{\bf{S}}_g+\\
+\frac{2t\alpha}{U}\,\frac{2-x^2}{1-x^2}\cos(\chi_{fg})\left(\left[{\bf{d}}_z\times{\bf{d}}_{fg}\right]\cdot \left[{\bf{S}}_f\times{\bf{S}}_{g}\right]\right)+\\
+\frac{4\alpha^2}{U}\,\frac{1}{1-x^2}\left(\left[{\bf{S}}_f\times{\bf{d}}_{fg}\right]_z\cdot \left[{\bf{S}}_g\times{\bf{d}}_{fg}\right]_z\right)\bigg],
\end{multline}
where ${\bf{d}}_z$ -- unit vector in $z$-direction (normal to the lattice direction), $\chi_{fg}=\frac{e}{c\hbar}\int_{{\bf{r}}=f}^{{\bf{r}}=g}{\bf{A}}\cdot d{\bf{r}}$ -- Peierls phase with $\bf{A}$ being the vector potential and $t$ is the value of $t_{fg}$ for nearest sites $f$ and $g$.

From expression (\ref{Heff2_tot}) (second line) it is seen the competition between AFM and FM exchange interactions with amplitudes $\sim t^2/U$ and $\sim \alpha^2/U$, respectively.
Besides, due to the spin-orbit interaction to types of chiral interactions emerge: DM type interactions (third line in (\ref{Heff2_tot})) and anisotropic exchange interaction depending on the of exchange bond direction (last line in (\ref{Heff2_tot})).

In the section \ref{sec2}, we considered the limit of $\alpha \ll t$ and hence the chiral interactions $\sim \alpha^2/U$ was discarded. However, analysis of its effect on the magnetic properties of the model (\ref{Hub_mod}) may be of interest. Of particular relevance to it is the observation that in the regime $t\approx\alpha$, the isotropic exchange interaction is significantly suppressed and thus, in this case, it is the chiral interaction that determines the properties of magnetic subsystem.

Performing similar calculations one obtains the expression for $\delta H_3$ in (\ref{H_eff_gen})
in terms of the matrices (\ref{XY_matrices}):
\begin{multline}
\delta\mathcal{H}_{3} = \frac{1}{2U^2}\sum_{\left[fgl\right]}Sp\Big[\hat{X}_f\hat{t}_{fg}\hat{Y}_g\hat{t}_{gl}\left(\hat{Y}_l-\hat{X}_l\right)\hat{t}_{lf}+h.c.\Big],\nonumber
\end{multline}
where by means of $\left[fgl\right]$ a triple of non equivalent site indexes is denoted.
Taking, at last, into account formulas (\ref{XY_S}) and leaving only the terms proportional to $t^3/U^2$ (discarding interactions $\sim t\alpha^2/U^2$, $\sim t^2\alpha/U^2$ and $\sim \alpha^3/U^2$ in the regime
$t\gg\alpha$) one comes to the desired expression for the chiral three-spin interaction:
\begin{multline}\label{H3_spinor}
\delta\mathcal{H}_{3}
\approx\frac{24\cdot t^3}{U^2}\sum_{\left[fgl\right]}\sin\left(\pi\Phi_{\Delta}\right)\left(\,{\bf{S}}_f\cdot\left[{\bf{S}}_g \times {\bf{S}}_l\right]\right).
\end{multline}
where $\Phi_{\Delta}$ is the magnetic field flux through a triangular placket (formed by three sites $f$, $g$ and $l$) written in units of the magnetic flux quantum $\phi_0=ch/2e$.

Note that in (\ref{H3_spinor}) the magnetic flux arose due to the Peierls substitution for the hopping integrals $t_{fg}$. This approximation assumes  $\Phi_{\Delta}\ll 1$.
In this case the value of $\delta\mathcal{H}_3$ is determined by the factor $24\cdot\pi t^3\Phi_\Delta/U^2$ which for competitiveness should be comparable to the DM interaction $\sim 4t\alpha/U$ (see section \ref{sec2}), that is:
$$\left(\frac{t}{U}\right)\left(6\pi\Phi_\Delta\right)\approx\left(\frac{\alpha}{t}\right).$$
It means that in the regime under consideration ($t/U\sim\alpha/t\ll 1$)
the value of the three-spin interaction (\ref{H3_spinor}) can be considered of the same order as DM interaction (third line in (\ref{Heff2_tot}) proportional to $ t\alpha/U)$ if the magnetic flax $\Phi_{\Delta}$ larger than $1/6\pi$.
Thus, the relevant interval for the flax $\Phi_\Delta$ is: $1/6\pi \lesssim \Phi_\Delta\ll 1$.

\section{Magnetic functionals of HOMS in an inhomogeneous magnetic field}\label{appendix_C}

Following the reference \cite{wang18} consider the problem of calculating energy functionals of the form
\begin{eqnarray}  \label{EF3}
{E} = E_{J}+E_{D}+E_{K}+E_{Z}+E_{A},
\end{eqnarray}
where
\begin{eqnarray}  \label{EJ3}
E_{J}&=& \frac{J}{2} \int_0^{\infty} \left[\left(\frac{d\Theta}{dr}\right)^2+\frac{n^2}{r^2}\sin^2\Theta\right]r\,dr,\\ \label{ED3}
E_{D}&=&\delta_{n,1}\cdot \frac{D}{\pi} \int_0^{\infty} \left[\frac{d\Theta}{dr}+\frac{\sin 2\Theta}{2r}\right]r\,dr, \\ \label{EK3}
E_{K}&=&\frac{K\,n}{2}\int_0^{\infty}{h}(r)~\sin \Theta ~ \frac{d\Theta}{dr}dr, \\ \label{EZ3}
E_{Z}&=& \frac{H}{2}\int_0^{\infty}  h(r)\,(1-\cos \Theta\,)\,r\,dr,\\ \label{EA3}
E_{A}&=&\frac{A}{2} \int_0^{\infty} \sin^2 \Theta ~ r\,dr. \label{EA3}
\end{eqnarray}
with a parametrizing function $\Theta(r,\,R,\,w)$ defined in  (\ref{sk_param}). In the case of a skyrmion with a narrow domain wall $R\gg w$, the integrands in (\ref{EJ3})-(\ref{EA3}) are noticeably different from zero only in the range $0<r<R+2w$ of the variable $r$. In this range the parametrizing function can be approximated by a function of the form:
\begin{eqnarray*}
\Theta(r,R,w) = 2\arctan(\,e^{\rho-t}\,),~~\rho = R/w,~~t = r/w.
\end{eqnarray*}
In this case the functions in the integrals (\ref{EJ3})-(\ref{EA3}) can be represented as:
\begin{eqnarray}\label{functions}
&~&\frac{d\,\Theta}{d\,r}=-\left(\frac{1}{w}\right)\frac{1}{\cosh{(t-\rho)}},~~\sin\,\Theta = \frac{1}{\cosh{(t-\rho)}},\nonumber\\
&~&\sin\,2\Theta=2~\frac{\tanh(t-\rho)}{\cosh(t-\rho)},~~~1-\cos\Theta = \frac{e^{\rho-t}}{\cosh(t-\rho)}. \nonumber\\
\end{eqnarray}
\textcolor{black}{Moreover, the some of integrands in Eq.\,(\ref{EJ3})-(\ref{EA3}) can be approximated by delta functions, sinse the functions
\begin{eqnarray}
&&\sin^2\Theta =\frac{1}{\cosh^2{(t-\rho)}}\to 2\,w\,\delta\left(\,R-r\,\right),\\
&&\left(\frac{d\Theta}{dr}\right)^2 \to \frac{2}{w}\,\delta\left(\,R - r\,\right)\,;~~\left(\frac{d\Theta}{dr}\right)\,\sin\Theta \to -2\,\delta\left(\,R - r\,\right)\,.\nonumber
\end{eqnarray}
localized around $r\approx R \gg 1$ and has finite value for $r=R$. So, we can use the mean value theorem for integrals over $r$ and consider these integrands as delta function.}

The function $\sin\,2\Theta$ in (\ref{functions}) significantly differ from zero for $t\approx \rho$, at the point $t=\rho$ is equal zero and has the different signs in the left and right neighborhoods of this point. The integral of this function can be neglected in the leading approximation. So, without Zeeman splitting the skyrmion energy can be approximated as
\begin{multline*}
E_{J}+E_{D}+E_{K}+E_{A} = \\
=J\left(\,\frac{R}{w} + \frac{w}{R}\,\right)-\delta_{n,1}\,D\,R+A\,R\,w - K\,n\,h(R)
\end{multline*}
and can be estimated for an arbitrary profile, $h(r)$, of axially symmetric magnetic field. For homogeneous fields the scalar chiral interaction shifts the energy of skyrmions by a value proportional to their topological charge.

\textcolor{black}{When taking into account the Zeeman contribution, Eq.\,(\ref{EZ3}) to the energy functional, we have to take into consideration that the function $1-\cos\Theta$ has finite values at $r \ll R$. Therefore, when considering the power-law profiles of the magnetic field, we should artificially set the field strength equal to zero in a small neighborhood of the origin:
\begin{eqnarray}\label{hvsr}
 h(r) = \left\{ \begin{array}{*{20}{c}}
0\,,~~~r < \delta r \ll 1~~~ \\
r^{\beta}\,,~~r \geq \delta r,~~\beta \in \mathbb{R} \end{array} \right.,\end{eqnarray}
This will allow us to regularize the integral Eq.\,(\ref{EZ3}) for $\beta < -1$. Moreover, such a choice of profile allows us to express such an integral using of the incomplete polylogarithm special function:}
\begin{multline*}\label{E_Z}
E_{Z}=H\,w^{\beta+2}\int_{\textcolor{black}{\delta r}}^{\infty}\frac{t^{\beta+1}dt}{1+e^{2(t-\rho)}}=\\
=-H\left(\frac{w}{2}\right)^{\beta+2}\Gamma_{\beta+2}\Li_{\beta+2}\left(-e^{2R/w}, \textcolor{black}{\delta r}\right)
\end{multline*}
where $\Gamma_{\beta+2}$ --- gamma function of argument $\beta +2$.  Hence, the energy of HOMS with topological charge $|Q|=n$ can be estimated as
\begin{eqnarray*}
E &=& J\left(\,\rho + n^2\rho^{-1}\,\right) - D\,\rho\,w\,\delta_{n,1} + A\,\rho\,w^2 +\nonumber\\
&+&\left(E^{(\beta)}_K + E^{(\beta)}_Z\right),\\
E^{(\beta)}_Z &=& -H\left(\frac{w}{2}\right)^{\beta + 2}\Gamma_{\beta+2}\,\Li_{\beta+2}\left(-e^{2\rho},\,\textcolor{black}{\delta r}\right);\nonumber\\
E^{(\beta)}_K &=& -K\,n \,\rho^{\beta}\,w^{\beta}.\nonumber
\end{eqnarray*}
\textcolor{black}{For the profiles with $\beta > -1$ it is convenient to put $\delta r = 0$. Then the incomplete polylogarithm become the polylogarithm function with a well-known asymptotic expansion for $\rho \gg 1$:}
\begin{eqnarray}
&~&\Li_{\beta+2}\left(-e^{2\rho} \right)=\\
&=&\sum_{2k \leq \beta + 2}(-1)^k\left(1- 2^{1-2k} \right)\frac{\left(2\pi\right)^{2k}}{2k\,!}\frac{B_{2k}}{\Gamma_{\beta+3-2k}}\left(2\rho\right)^{\beta+2-2k}, \nonumber
\end{eqnarray}
where $B_{2k}$--- Bernoulli numbers.

\section{Analytical minimization of energy functional}\label{appendix_D}

Consider the question of solving Eq.(\ref{w5_sms_ddm}) in the case $A=0$. It is convenient to find the solution $w(z)$ as an expansion in the free coefficient, $z$ of modified Eq.(\ref{w5_sms_ddm}) with the use the formula of Ref.\,\cite{aizenberg79}
\begin{eqnarray}
\label{yu}
a = f
\end{eqnarray}

%\begin{align}
%w = z^{\frac{1}{k}}+\sum_{\alpha \geq %2}\frac{z^{\frac{\alpha}{k}}}%{\alpha\,!}\,\mathcal{D}^{\alpha}\left.\left[\,\frac{w}{\Psi^{\frac{\alpha+1}{k}}}\,\left(w\,\Psi^{\frac{1}{k}}\right)_{w}^{'}\,\right]\right|_{w=0},
%\end{align}\label{yujakov}
where it is assumed that
\begin{eqnarray}\label{yujakov_condition}
w^k\,\Psi(w)=z\,;~~~\Psi(0)=1,
\end{eqnarray}
and $\mathcal{D}^{\alpha}$ denotes the $\alpha$-order derivative by $w$. For the Eq.(\ref{w5_sms_ddm}) we have
\begin{eqnarray}\label{Psi_AB}
    &&k=2,~~\Psi(w)=1+B\,w+A\,w^3,\\
    &&z = \frac{K\,n}{{H}(n^2+c^2)},~~
    A = \frac{-2{H}\,c^4}{3J\,(n^2+c^2)},~~B = \frac{2\,K\,n\,c^2}{3J(n^2+c^2)}.\nonumber
\end{eqnarray}
Then, making calculations according to (\ref{yu}), we obtain
\begin{eqnarray}\label{w_sol_gen}
w = \sqrt{z}-\frac{B}{2}\left(\sqrt{z}\right)^2 + \frac{5B^2}{8}\left(\sqrt{z}\right)^3+\frac{1}{2}\sum_{\alpha=4}^{\infty}\left(-\sqrt{z}\right)^{\alpha}\cdot \nonumber \\
\cdot\Biggr\{-2\sum_{i=0}^{\left[\frac{\alpha-1}{3}\right]}\frac{B^{\alpha-1-3i}A^{i}}{\left(\alpha-1-3i\right)!\,i!}\left(\frac{\alpha}{2}+1\right)\dots\left(\frac{3\alpha}{2}-1-2i\right)+\nonumber \\
+3B\sum_{i=0}^{\left[\frac{\alpha-2}{3}\right]}\frac{B^{\alpha-2-3i}A^{i}}{\left(\alpha-2-3i\right)!\,i!}\left(\frac{\alpha}{2}+1\right)\dots\left(\frac{3\alpha}{2}-2-2i\right)+\nonumber \\
+5A\sum_{i=0}^{\left[\frac{\alpha-4}{3}\right]}\frac{B^{\alpha-4-3i}A^{i}}{\left(\alpha-4-3i\right)!\,i!}\left(\frac{\alpha}{2}+1\right)\dots\left(\frac{3\alpha}{2}-4-2i\right)\Biggr\},\nonumber\\
\end{eqnarray}
where $\left[\,c\,\right]$ is the rounding $c$. The first three terms on the right-hand side of Eq.\,(\ref{w_sol_gen}) fulfil an approximate solution (\ref{Rw_an_H}) for a higher-order magnetic skyrmion sizes

Let us note the key steps for obtaining (\ref{w_sol_gen}). The derivative $\mathcal{D}^{\alpha}(F\cdot G_{\alpha})$ has been calculated with the use the Newton's binomial
$$\left.\frac{\mathcal{D}^{\alpha}}{\alpha!}\left(F\cdot G_{\alpha}\right)\right|_{w=0} = \sum_{k=0}^{\alpha}\frac{1}{k!(\alpha-k)!}F^{(k)}(0)\cdot G_{\alpha}^{(\alpha-k)}(0),$$
where functions
$$F(w) = 2w + 3Bw^2 + 5Aw^4;~G_{\alpha}(w)=\left(1 + Bz + Az^3\right)^{-\frac{\alpha+2}{2}}$$
arose naturally from (\ref{yu}) and (\ref{Psi_AB}). Calculation of the $n$-order derivative of the function $G^{(n)}_{\beta}=g^{-\beta}$, where $g = 1 + Bz + Az^3$ and $\beta = (\alpha+2)/2$, was carried out using the Faà di Bruno's formula \cite{kaufmann75}
\begin{eqnarray}\label{faa_di_bruno}
&&G_{\beta}^{(n)}(w)
=\sum^{n}_{k=1}\sum_{\genfrac\{\}{0pt}{2}{k=\alpha_1+\ldots+\alpha_n}{n=1\alpha_1+\ldots+n\alpha_n}}\\
&&\frac{n\,!}{\alpha_1!\ldots\alpha_n!}\cdot
\left(\,g^{-\beta}\,\right)^{(k)}\,\left(\,\frac{g^{(1)}}{{1\,!}}\,\right)^{\alpha_1}\ldots\left(\,\frac{g^{(n)}}{{n\,!}}\,\right)^{\alpha_n}.\nonumber
\end{eqnarray}
Since we find $G_{\beta}^{(n)}(0)$ and the $g(w)$ has the property
$$g^{(0)}(0)=1,~~g^{(1)}(0)=B,~~g^{(3)}(0)=6A,$$
thus the summation over combinatorial sets in the Eq.\,(\ref{faa_di_bruno}) is bounded by two conditions:
$k = \alpha_1 + \alpha_3,~~n = \alpha_1 + 3\alpha_3.$
With these conditions taken into account, the triple summation over $k$, $\alpha_1$, and $\alpha_3$ was reduced to a single one, resulting in the answer (\ref{w_sol_gen}).

Note that the above formula (\ref{yu}) can be generalized used to solve a system of $N\times N$ algebraic equations of type (\ref{yujakov_condition}). Then the variables $w_{i}$ and free terms $z_{i}$ become component-dependent. For magnetic skyrmions such systems can arise when describing the complex nontrivial magnetic structures, such as magnetic skyrmionium, skyrmion bags etc., characterized by several radii, $R_i$, and domain wall widths $w_i$. Wherein, it is important to note that the generalization of the Eq.\,(\ref{yu}) is valid if the condition (\ref{yujakov_condition}) is satisfied.

In general, for systems of $n$ algebraic equations
\begin{equation}\label{syst}
\sum_{\lambda \in A^{(i)}}a^{(i)}_{\lambda}\,y^{\lambda},~~~i=1,\ldots,n\,,
\end{equation}
with variables $y=\left(y_1\,,\ldots,\,y_n\right)$ and coefficients $a^{(i)}_{\lambda}$, where $A^{(i)}\in \mathbb{Z}^n$ -- fixed finite subsets of an integer lattice, $y^{\lambda}=y_1^{\lambda_1}\ldots y_n^{\lambda_n}$ there is a formula for the solution (\ref{syst}) in the form of hypergeometric series \cite{sadykov14}. However, in the case of small $n$, it may be more practical to eliminate multiple variables by the method described in Refs.\cite{arzhantsev03, bykov91} to obtain one equation. The latter can be solved by the methods of Refs.\,\cite{birkeland03, mellin21}. Such approaches assume the presence of a pair of terms with coefficients that are an order of magnitude greater than the coefficients for the other terms. If there is one largest coefficient (\ref{yu}) can be applied.

\bibliography{homs}

\end{document}